\documentclass[prd,aps,eqsecnum,floatfix,nofootinbib,preprint,tightenlines]{revtex4}

\usepackage[utf8]{inputenc}
\usepackage{latexsym}
\usepackage{graphicx}
\usepackage[dvipsnames]{xcolor}

\def\bibi{\bibitem}
\def\floatcaption#1#2{ \caption{#2 \label{#1}} }

\let\inodot=\i

\def\a{\alpha}
\def\b{\beta}
\def\c{\chi}
\def\d{\delta}
\def\e{\epsilon}                
\def\g{\gamma}

\def\i{\iota}

\def\m{\mu}

\def\p{\pi}                     
\def\r{\rho}                    
\def\s{\sigma}                  

\def\D{\Delta}

\def\P{\Pi}

\def\S{\Sigma}

\def\co{{\cal O}}

\def\cbo{{\,\raise-.15ex\Sc [\,}}                       
\def\ltap{\raisebox{-.4ex}{\rlap{$\sim$}} \raisebox{.4ex}{$<$}}   

\def\svev#1{\left\langle #1\right\rangle}       

\def\ddt#1{{\buildrel {\hbox{\LARGE .\kern-2pt.}} \over {#1}}}

\def\ie{\mbox{\it i.e.}}

\def\etc{\mbox{\it etc.}}

\def\half{{1\over 2}}

\def\ttl#1{{\it #1}}

\def\seef{{\it cf.}}

\def\amuHVP{{{$a_\m^{\rm HVP}$}}}

\begin{document}

\begin{center}
\vspace*{5mm}
\begin{boldmath}
{\large\bf The muon anomalous magnetic moment with staggered fermions:\\ is the lattice spacing small enough?}
\end{boldmath}
\\[10mm]
Christopher Aubin,$^a$ Thomas Blum,$^b$ Maarten Golterman$^{c,d}$ and 
Santiago Peris$^d$
\\[8mm]
\null$^a$Department of Physics \& Engineering Physics\\ Fordham University, Bronx,
New York, NY 10458, USA\\
\null$^b$Physics Department\\
University of Connecticut, Storrs, CT 06269, USA\\
\null$^c$Department of Physics and Astronomy, San Francisco State University\\
San Francisco, CA 94132, USA\\
\null$^d$Department of Physics and IFAE-BIST, Universitat Aut\`onoma de Barcelona\\
E-08193 Bellaterra, Barcelona, Spain
\\[6mm]

{ABSTRACT}
\end{center}

We extend our previous work on the light-quark connected part, $a_\m^{\rm HVP,lqc}$, of the leading order hadronic-vacuum-polarization (HVP) contribution to the muon anomalous magnetic moment 
$a_\m$, using staggered fermions, in several directions. We have collected more statistics on ensembles with lattice spacings of $0.06$, $0.09$ and $0.12$~fm, and we added two new
ensembles, both with lattice spacing $0.15$~fm, but with different volumes.   The increased statistics allow us to reduce statistical errors on $a_\m^{\rm HVP,lqc}$ and related
window quantities significantly.  We also calculate the current-current correlator from which $a_\m^{\rm HVP,lqc}$ is obtained to next-to-next-to-leading order (NNLO)
in staggered chiral perturbation theory, so that we can correct lattice values for $a_\m^{\rm HVP,lqc}$ to NNLO for finite-volume, pion-mass mistuning and taste-breaking effects.  We discuss the applicability of NNLO chiral perturbation theory to $a_\m^{\rm HVP,lqc}$ and to the window quantities, emphasizing that it provides a systematic EFT approach to $a_\m^{\rm HVP,lqc}$, but not to short- or intermediate-distance window quantities.  This makes it difficult to assess systematic errors on the standard
intermediate-distance window quantity that is now widely considered in the literature.  In view of this, we investigate a longer-distance window, for which EFT methods should be more reliable.   Our most important conclusion is
that, especially for staggered fermions, new high-statistics computations at lattice spacings smaller than $0.06$~fm are indispensable.
\\[2mm]

\begin{quotation}
\end{quotation}

\section{\label{intro} Introduction}

The recent confirmation \cite{FNL} of the experimental value \cite{BNL} for the anomalous magnetic moment of the muon, $a_\m$, has now sharpened the discrepancy with the Standard-Model (SM) estimate \cite{whitepaper} to 4.2 standard deviations ($\s$).  As is well known, the largest part of the uncertainty in the 
SM estimate comes from the hadronic corrections, which appear first at 
order $\a^2$ ($\a$ is the fine-structure constant) through the contribution from the hadronic vacuum polarization, \amuHVP, and at order $\a^3$ through the hadronic light-by-light contribution $a_\mu^{\rm HLbL}$.

 The SM estimate is based on a data-driven evaluation of \amuHVP\ through dispersive methods, while both data-driven and lattice methods contribute to the 
current best value for $a_\mu^{\rm HLbL}$; for the latter, data-driven and lattice estimates are in good agreement \cite{whitepaper,MainzHLbL}.  For \amuHVP\ the situation is more complicated:  while the uncertainties of most determinations based on lattice QCD do not resolve at present the discrepancy between the experimental and SM values, one collaboration \cite{BMW20} finds a value for \amuHVP\  leading to an SM estimate about $1.5\s$ below the experimental value, and about $2.1\s$ above the estimate based on the dispersive value.

Lattice-based determinations are afflicted by a number of systematic errors, with finite-volume (FV) corrections, continuum extrapolation and scale setting among the most important of these.  All lattice collaborations compute the various contributions to \amuHVP\ using\footnote{Or aiming for.} at least three lattice spacings, allowing, in principle, for a continuum extrapolation.  In contrast, estimating FV corrections purely by numerical extrapolation to the infinite-volume limit is too costly, and effective-field-theory (EFT) methods and models play an important role in estimating these corrections.  As, in general, ensembles at different lattice spacings do not have the same spatial volume, FV corrections have to be applied at each lattice spacing, before extrapolation to the continuum limit is attempted (also, even if physical pion masses are used, there are small mistunings to be corrected).  Good control over these systematic errors is particularly important for the light-quark connected part of \amuHVP, as it contributes about 90\% to the total.

In the time-momentum representation \cite{BM11}, \amuHVP\ is obtained from 
\begin{equation}
\label{amuHVP}
a_\m^{\rm HVP}=2\int_0^\infty dt\ w(t) \,C(t)\ ,
\end{equation}
with
\begin{equation}
\label{Ct}
C(t)=\frac{1}{3}\sum_i\int d^3 x\svev{j_i({\vec x},t)j_i({\vec 0},0)}\ ,
\end{equation}
where $j_i=\frac{2}{3}\bar{u}\g_i u+\dots$ are the spatial components of the hadronic contribution to the electromagnetic current,\footnote{We use conserved currents on the lattice.} and the weight $w(t)$ is defined by
\begin{eqnarray}
\label{weight}
w(t)&=&4\a^2\int_0^\infty dQ^2\left(\frac{\cos{(Qt)}-1}{Q^2}+\frac{1}{2}\, t^2\right)f(Q)\ ,\\
f(Q)&=&\frac{m_\m^2 Q^2 Z^3(Q)(1-Q^2 Z(Q))}{1+m_\m^2 Q^2 Z^2(Q)}\ ,\qquad
Z(Q)=\frac{\sqrt{Q^4+4Q^2 m_\m^2}-Q^2}{2m_\m^2 Q^2}\nonumber
\end{eqnarray}
(for a detailed discussion of this weight, see Ref.~\cite{Mainz17}; for the momentum representation in momentum space obtained by integrating over $t$, 
see Refs.~\cite{LR,TB}, \seef\ Eq.~(\ref{amuHVPmom}) below).  We have used that $C(t)$ and $w(t)$ are even functions of $t$.   

A ``window'' $W(t)$
can be introduced to define ``window quantities'' \cite{RBC18}
\begin{equation}
\label{Wquantity}
a_\mu^{\rm W}(t_0,t_1,\D)=2\int_0^\infty dt\ W(t;t_0,t_1,\D)\,w(t) \,C(t)\ ,
\end{equation}
with $W(t;t_0,t_1,\D)$ defined by
\begin{equation}
\label{windowdef}
W(t;t_0,t_1;\D)=\half\left(\tanh\left(\frac{t-t_0}{\D}\right)-\tanh\left(\frac{t-t_1}{\D}\right)\right)\ .
\end{equation}
This is a step function equal to one for $t_0\,\ltap\, t\,\ltap\, t_1$ and zero outside this interval, which transitions smoothly over a range $\D$.  These 
window quantities single out the contribution of a particular region in $t$ to the total \amuHVP.

It has become standard to compute $a_\m^{\rm W}(0.4,1;0.15)$ (with values
of the arguments in fm), since this intermediate-distance window can be computed with a
significantly higher precision than \amuHVP\ itself, and it thus provides a good benchmark for comparison between different lattice computations.

\begin{figure}[t!]
\vspace*{4ex}
\begin{center}
\includegraphics*[width=11cm]{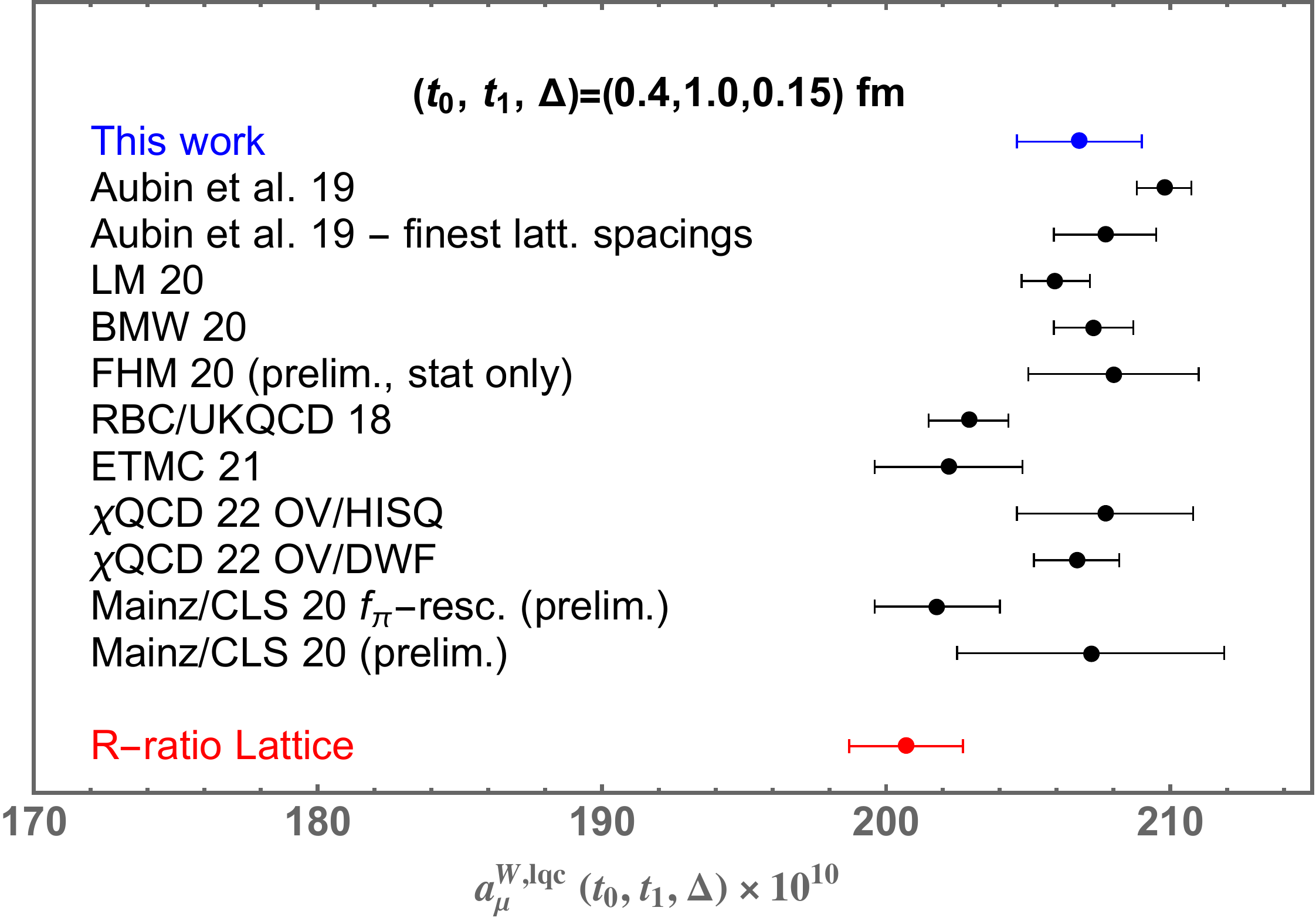}
\end{center}
\begin{quotation}
\floatcaption{fig:window}%
{{\it The isospin-symmetric, light-quark connected contribution to $a_\m^{\rm W}(0.4,1.0;0.15)$ from Aubin {\it et al.} \cite{ABGP19}, LM~20 \cite{LM20}, BMW~20 \cite{BMW20}, FHM~20 \cite{FHM20}, RBC/UKQCD~18 \cite{RBC18}, ETMC~21 \cite{ETMC20}, $\c$QCD~22 \cite{chiQCD22} and Mainz/CLS~20 \cite{Mainz20}. The blue point represents the value we
obtain in Sec.~\ref{standard} below.
The red point is obtained by using lattice results to convert \amuHVP\ from the dispersive approach to its isospin-symmetric, light-quark connected part (by C.~Lehner, using data from 
Ref.~\cite{KNT18}).}}
\end{quotation}
\vspace*{-4ex}
\end{figure}
Recent results for the light-quark connected contribution to 
$a_\m^{{\rm W}}(0.4,1.0;0.15)$ are summarized in Fig.~\ref{fig:window}.   The top five results are obtained using staggered fermions,\footnote{With all except BMW~20 using subsets of MILC-HISQ ensembles \cite{MILC}, so that the Aubin {\it et al.}~19, LM~20, FHM~20 and this work's results are to some extent correlated.  Our 48II ensemble (\seef\ Table~\ref{tab:ensembles} below) is a CalLat ensemble \cite{CalLat}.} while the bottom four results are obtained using Wilson-like
fermions (domain-wall, twisted-mass, and clover fermions, respectively). Besides discretization of the Dirac operator, the calculations differ in numbers of sea quark flavors, 2+1 $vs.$ 2+1+1, and whether conserved or local currents are used, which affects the path to the continuum limit.
The point denoted with ``R-ratio'' is obtained by correcting the dispersive
value of $a_\m^{\rm W}$ by subtracting lattice window results for the strange-quark, disconnected, \etc, parts from it.
The staggered results are not in agreement with the R-ratio value, while the Wilson-like results are
(except the very recent results from Ref.~\cite{chiQCD22}).   Clearly, the spread in these results needs to be understood in order to gain a better understanding of the discrepancy 
between lattice and dispersive values for \amuHVP.

In this paper, our aim is to 
update our earlier staggered results \cite{ABGP19} for the light-quark connected
contribution $a_\mu^{\rm HVP,lqc}$ to \amuHVP, using our new results to investigate the continuum limit, as well as the use of various methods to compute FV corrections.  
While we will also discuss the full quantity $a_\mu^{\rm HVP,lqc}$, our main focus will be on 
window quantities $a_\mu^{\rm W,lqc}$, using our results to study possible causes of the discrepancies visible in Fig.~\ref{fig:window}.  

In the case of staggered fermions, the breaking of taste symmetry at non-zero lattice spacing plays a special role.\footnote{For reviews of staggered fermions and taste breaking, see Refs.~\cite{MGLH,MILCreview}.}  New operators (proportional to powers of the lattice spacing $a$, starting with $a^2$) appear in the low-energy EFT describing the physics of the Nambu--Goldstone
bosons \cite{LS,AB}, leading to ``taste splittings'' \cite{MGmesons}
in their mass spectrum.  While this is an $\co(a^2)$ effect that disappears in the continuum limit, it has generically been found to be advantageous to ``correct'' for these taste splittings already at non-zero lattice spacing,
using staggered chiral perturbation theory (SChPT).   Since roughly half of the lattice effort aimed at \amuHVP\ employs staggered fermions, it is important to investigate the special role of taste breaking in taking the continuum limit.

Two different approaches have been used in the computation of FV effects, and to correct for taste breaking and pion-mass mistuning.  One approach is based on chiral perturbation theory (ChPT), the low-energy theory for QCD with light quarks.  This approach is based on a perturbative expansion
and power counting, and is systematically improvable by the calculation of higher orders in the expansion.  In Ref.~\cite{ABGPEFT} we showed that ChPT can be applied to \amuHVP\ systematically, and we applied this observation specifically to the calculation of FV effects.  Here we will extend this to a quantitative estimate for the approximate maximum value of the pion masses for which next-to-next-to-leading-order (NNLO) ChPT can be trusted.  We will also compute FV and taste-breaking effects to NNLO in SChPT.\footnote{In Ref.~\cite{ABGP19} staggered effects were only included to NLO while NNLO
FV effects were computed in continuum ChPT; Ref.~\cite{BMW20} was the first to extend their inclusion to NNLO.}   The key 
observation of Ref.~\cite{ABGPEFT} is that when considering \amuHVP, the appropriate EFT is an EFT for pions, muons and photons, thus extending the usual definition of ChPT to include QED effects (to leading order in $\a$, in our
case) due to the electromagnetic interactions of pions with the 
muon.  It was argued that this extended EFT framework allows for the introduction of all necessary counter terms in the EFT to make the chiral expansion systematically applicable to \amuHVP.

The other approach has relied on the use of models to understand these systematic effects.  In particular, $a_\m^{\rm W}(0.4,1.0;0.15)$ is a 
quantity defined at scales at which ChPT is not expected to work, as for example demonstrated in Ref.~\cite{BMW20}.  Applying ChPT to compute FV and
taste-breaking corrections for $a_\m^{\rm W}(0.4,1.0;0.15)$ nevertheless
thus degrades ChPT to a model, and one would expect better models to
exist, by including the physics of the $\r$ meson more completely than it is
at NNLO in ChPT.  Here we will pay particular attention to the ``SRHO'' model,
developed specifically for the case of staggered fermions in Ref.~\cite{HPQCD16} and applied further in Refs.~\cite{BMW20,FHM19}.
The disadvantage is, of course, that even if a model may work very well for a certain range of parameters, extrapolations are less well controlled, and there is no path to systematic improvement of the model.

This discussion points at a disadvantage of the window quantity
$a_\m^{\rm W}(0.4,1.0;0.15)$, which is now computed by many collaborations for the sake of comparison between different methods.   If reliable extrapolations to the continuum limit and infinite volume could be obtained directly from the lattice data, this comparison would be unambiguously 
meaningful, and thus very useful, as $a_\m^{\rm W}(0.4,1.0;0.15)$ is a physical quantity, and can
be computed with very small statistical errors.   However, in practice, 
these limits cannot be easily obtained directly from the data, and we
need EFT or model methods to compute FV corrections, and, in the case
of staggered fermions, taste-breaking corrections to improve the continuum limit.   But, for $a_\m^{\rm W}(0.4,1.0;0.15)$ no reliable EFT-based method is available, and we thus need to resort to models.  One might expect that FV corrections and taste-breaking corrections for $a_\m^{\rm W}(0.4,1.0;0.15)$
are very small, but with the small statistical errors, one needs quantitative information on how small.  Using both ChPT and the SRHO model, we will see that, while they are indeed small, it is unlikely that these corrections can be safely neglected, and indeed the systematics of these corrections dominate the total error.   This calls the usefulness of this window quantity into question,
at least with the current state of the art.   We will therefore explore
another, longer-distance, window quantity, $a_\m^{\rm W}(1.5,1.9;0.15)$, which we will 
show to be more accessible to ChPT.  The disadvantage is that lattice 
computations of this window quantity will have larger statistical errors,
and it remains to be seen which window quantities will turn out to be
optimal for comparison between different lattice results.   We will
refer to these two windows as W1 and W2, with 
\begin{equation}
\label{windowdefs}
a_\m^{\rm W1}=a_\m^{\rm W}(0.4,1.0;0.15)\ ,\qquad a_\m^{\rm W2}=a_\m^{\rm W}(1.5,1.9;0.15)\ .  
\end{equation}
As 
always in this paper, we will consider the light-quark connected contributions $a_\m^{\rm W1,lqc}$ and $a_\m^{\rm W2,lqc}$ to $a_\m^{\rm W1}$ and $a_\m^{\rm W2}$.

This paper is organized as follows.   In Sec.~\ref{data}, we present our new
data from the lattice, and provide brief details about the computation.
In Sec.~\ref{ChPT}, we revisit ChPT, presenting our extension of
SChPT to NNLO, we give an estimate for the maximal pion mass at which NNLO ChPT can be expected to be quantitatively reliable, and we compare taste breaking on our ensembles with SChPT.   Our results for $a_\m^{\rm HVP,lqc}$,
including FV and taste-breaking corrections are then presented in Sec.~\ref{full}, and we convert our continuum extrapolation of $a_\m^{\rm HVP,lqc}$ into a 
value for \amuHVP\ using results for other contributions from the literature in Sec.~\ref{fullamu}.  We then present our results for light-quark connected window quantities in Sec.~\ref{windows}.   In both Secs.\ \ref{full} and \ref{windows}, the extrapolation to the continuum limit is discussed.
Section~\ref{concl} contains our conclusions.   An appendix briefly details our implementation of the SRHO model.

\section{\label{data} Lattice computation}
We summarize our ensembles in Table~\ref{tab:ensembles}.
\begin{table}[t]
\begin{center}
\begin{tabular}{|c|c|c|c|c|c|c|c|c|} 
\hline
label & $a$ (fm) & $L^3\times T$ & $m_\p$ (MeV) & $m_S$ (MeV) & $m_\p L$ & \#configs & sep. & \#low modes \\
\hline
96 & 0.05684 & $96^3\times 192$ & 134.3 & 153 & 3.71 & 77 & 60 & 8000\\
64 & 0.08787 & $64^3\times 96$ & 129.5 & 212 & 3.69 & 78 & 100 & 8000\\
48I & 0.12121 & $48^3\times 64$ & 132.7 & 326 & 3.91 & 32 & 100 & 8000\\
32 & 0.15148 & $32^3\times 48$ & 133.0 & 418 & 3.27 & 48 & 40 & 8000 \\
48II & 0.15099 & $48^3\times 64$ & 134.3 & 418 & 4.93 & 40 & 100 & 8000\\
\hline
\end{tabular}
\end{center}
\vspace*{-3ex}
\begin{quotation}
\floatcaption{tab:ensembles}{{\it Parameters defining the lattice ensembles.   Columns contain a label to refer to the ensemble, the lattice spacing $a$, the spatial volume $L^3$ times the temporal direction $T$ (in lattice units), the Nambu--Goldstone pion mass $m_\p$, the maximum pion mass $m_S$ in the pion taste multiplet (\seef\ Sec.~\ref{taste}), $m_\p L$, the number of configurations in the ensemble, the separation between measurements ({\rm ``sep.''}), and the number of low-mode eigenvectors.}}
\end{quotation}
\vspace*{-4.5ex}
\end{table}
In comparison with Ref.~\cite{ABGP19}, we extended our data set in several directions.   We employed more configurations and low-mode eigenvectors for the 
first three ensembles shown in the table, and we added two ensembles at a coarser lattice spacing, with different volumes.  We summarize our
methodology in Sec.~\ref{method}, and we present the results in Sec.~\ref{results}.

\subsection{\label{method} Methodology}

The lattice computation here is an extension of the previous simulations of Ref.~\cite{ABGP19}. First, as can be seen in Table~\ref{tab:ensembles}, we have run on more configurations of the finest two ensembles (96 and 64) and we have increased
the number of low modes used (from 4000 to 8000 low modes on 96 and from 6000 to 8000 on the 48 and 64 ensembles) in order to improve the statistics. We increased the number of trajectories separating measurements 
significantly on all three ensembles, in particular on the $64$ ensemble.  Additionally we have included two coarse ensembles with $a\approx 0.15$~fm at two volumes, $32^3$ and $48^3$ (where the  coarse $48^3$ ensemble is labeled as 48II to distinguish it from the $a\approx 0.12$ fm $48^3$ ensemble, labeled as 48I), both to include an additional lattice spacing and to examine  finite volume effects explicitly. All ensembles are near the physical pion mass.

We continued using the noise-reduction techniques combining full-volume low-mode averaging \cite{LMA0,LMA1,LMA2,LMA3} and all-mode averaging developed by the RBC and UKQCD collaborations \cite{RBC18,BlumLMAAMA,CovAA}. We omit the specific details as they are the same as in Ref.~\cite{ABGP19}.

Finally, as in our previous work, we implement the bounding method of Refs.~\cite{RBC18,BMW18} to further reduce statistical errors on the extraction of $a_\m^{\rm HVP,lqc}$. 
Here we use that the correlator of Eq.~(\ref{Ct}) has a lower bound of 0 for $t>t_b$ and an upper bound of
$C(t_b) e^{-E_0(t-t_b)}$,
where $E_0 = 2\sqrt{m_\pi^2 + (2\pi/L)^2}$ is the lowest (two-pion) energy state in the vector channel.
For sufficiently large $t_b$, the two bounds
overlap to give a more precise
result for $a_\m^{\rm HVP,lqc}$ than we would obtain by summing over the long-distance tail. 

\subsection{\label{results} Results}
The simulations on the ensembles listed in Table~\ref{tab:ensembles} provide us with measurements of the correlator $C(t)$ defined in Eq.~(\ref{Ct}), where we used conserved currents.   We then use trapezoidal integration, defining
\begin{eqnarray}
\label{amut}
a_\m(t)&
\equiv & 2a\sum_{n=1}^{[t/a]-1}w(na)C(na)+aw([t/a]a)C([t/a]a)\\
&\to & 2\int_0^t dt'\,w(t')C(t')\qquad \mbox{for}\ a\to 0\ .\nonumber
\end{eqnarray}
On the lattice, the maximum value of $t$ in Eq.~(\ref{amut}) is $t/a=T$,
with $T$ given in Table~\ref{tab:ensembles} for the five ensembles.
In the limits $a\to 0$ and $T\to\infty$, this yields \amuHVP, as defined in Eq.~(\ref{amuHVP}), but at non-zero $a$ there is an $\mathcal{O}(a^2)$ correction. For window quantities, the window function $W(t;t_0,t_1,\D)$ is inserted with appropriate choices for the window parameters $t_0$, $t_1$ and $\D$. 

In Fig.~\ref{fig:results} we show $a_\m(T;t_b)$ as a function of $t_b$ (in fm).   Here $t_b$ is the
value of $t$ in Eq.~(\ref{amut}) in which we switch from the lattice correlator $C(t)$ to the upper bound
(blue points) or lower bound (orange points) replacing $C(t)$ from $t=t_b$ to $t=aT$ by the upper or lower
bound, following the
bounding method of Ref.~\cite{RBC18}, \seef\ Sec.~\ref{method}.   The shaded bar indicates the $t_b$ region from which we obtain
the values for $a_\m^{\rm HVP,lqc}$ shown in Table~\ref{tab:results}; we did not use the bounding
method for the two window quantities also shown in the table as they depend much less on the long-time
tail of $C(t)$.
Errors are statistical only.
The lattice spacing is set using $w_0=0.1714(14)$~fm \cite{MILCw0}.\footnote{Even though \amuHVP\ is a dimensionless quantity, one needs to set the scale of the hadronic physics relative to the muon mass.}     As expected, the statistical errors for the longer-distance window W2 fall between those for $a_\mu^{\rm HVP,lqc}$ and those for the 
shorter-distance window W1.  There is a strong dependence on the lattice spacing, and, for $a_\mu^{\rm HVP,lqc}$, a clear dependence on volume, shown by the difference of the value for the last two ensembles in the table, which differ only by volume. The volume dependence of the $a=0.15$~fm ensembles is not visible in the two window
quantities.  Here one should keep in mind that taste-breaking effects are large for $a=0.15$~fm, and FV effects would be much larger were they not ``masked'' by the taste splittings.   All these systematic effects will be discussed in the following sections.

We can compare the values of Table~\ref{tab:results} with those of Tables II
and IV of Ref.~\cite{ABGP19}, for the first three ensembles.\footnote{We did not consider the window $\hat{w}$ of Ref.~\cite{ABGP19}, as we always found that it leads to a stronger dependence on the lattice spacing.}  
Different configurations were used in this paper, making our new results independent of the results of Ref.~\cite{ABGP19}.
 For the 96 and 64 ensembles, there is a significant reduction in the statistical error for $a_\mu^{\rm HVP,lqc}$ and $a_\m^{\rm W1,lqc}$ (the W2 window was not considered in Ref.~\cite{ABGP19}).   We notice that there is a mild tension
 between the 96- and 64-ensemble results for $a_\m^{\rm W1,lqc}$, up to about $1.4\sigma$.  This would in principle
 allow us to combine the results of Ref.~\cite{ABGP19} with our new results, but we chose not to do this.   First, 
 it is not excluded that the tension is caused by the smaller separation between measurements (notably for the 64
 ensemble) in Ref.~\cite{ABGP19}.  Second, including also the systematic errors to be discussed below, combining the
 two data sets does not lead to a significant reduction in errors compared to those obtained with only the new 
 data set.

\begin{figure}[t!]
\vspace*{4ex}
\begin{center}
\includegraphics*[width=7cm]{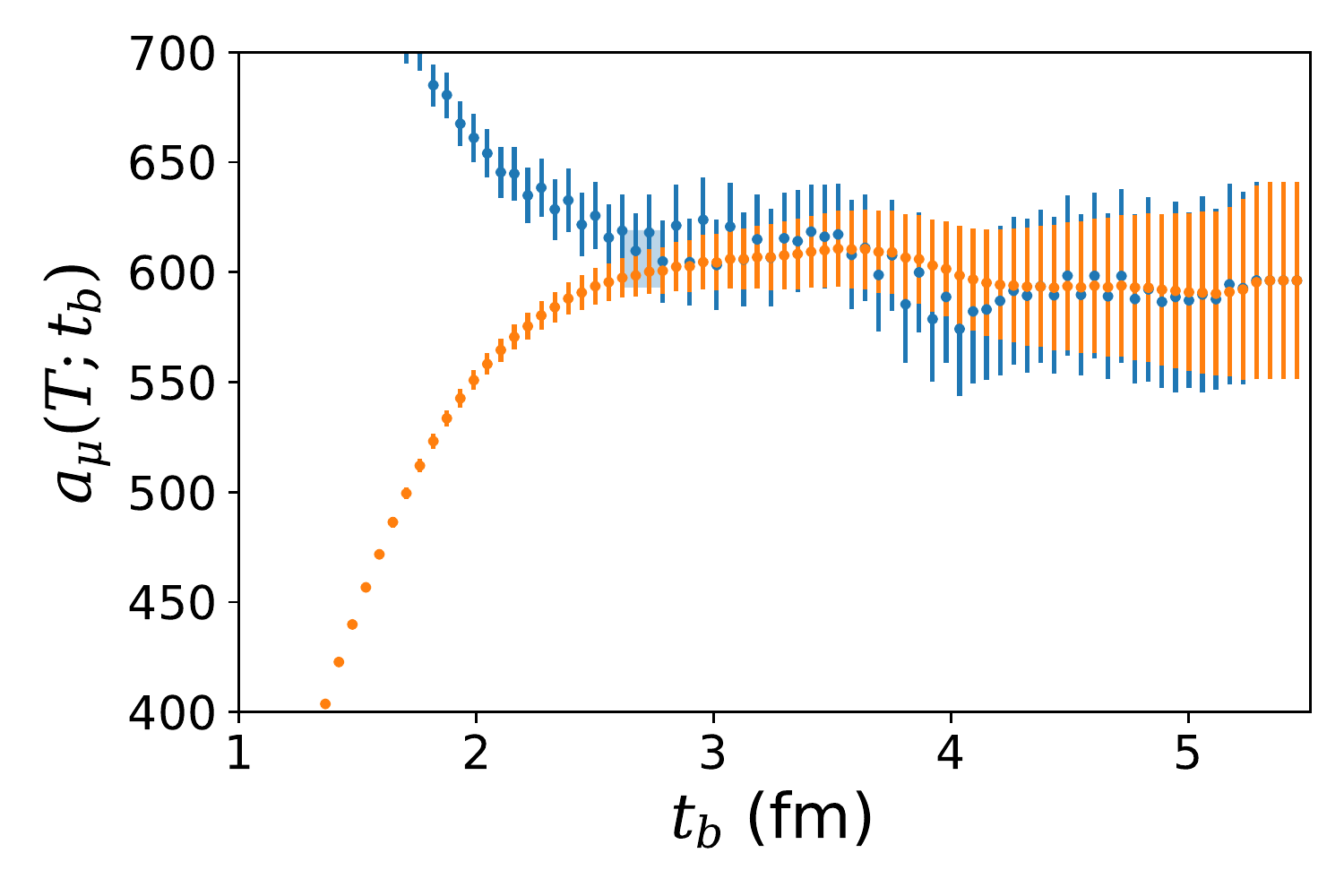}
\hspace{0.5cm}
\includegraphics*[width=7cm]{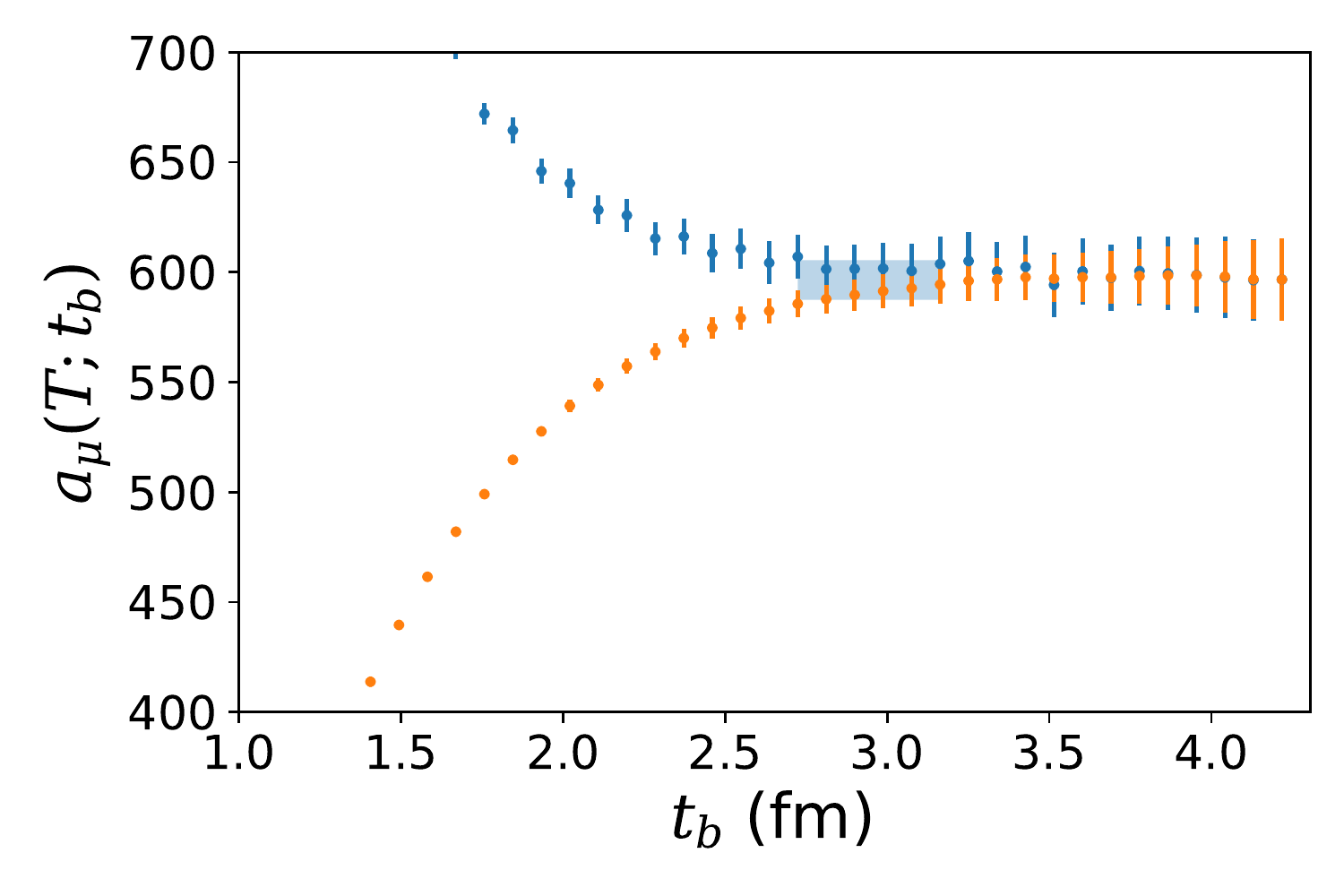}
\vspace{0.5cm}
\includegraphics*[width=7cm]{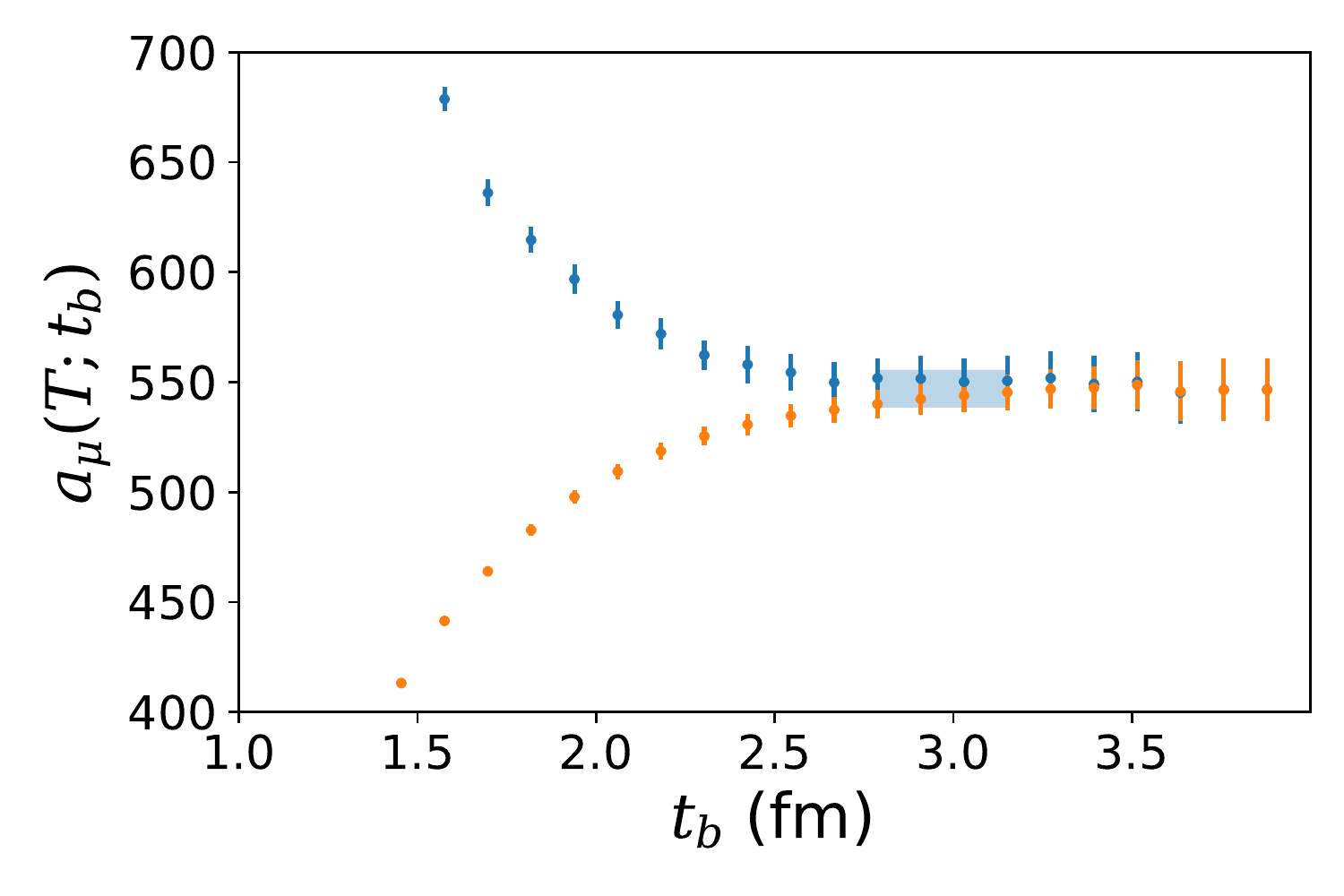}
\hspace{0.5cm}
\includegraphics*[width=7cm]{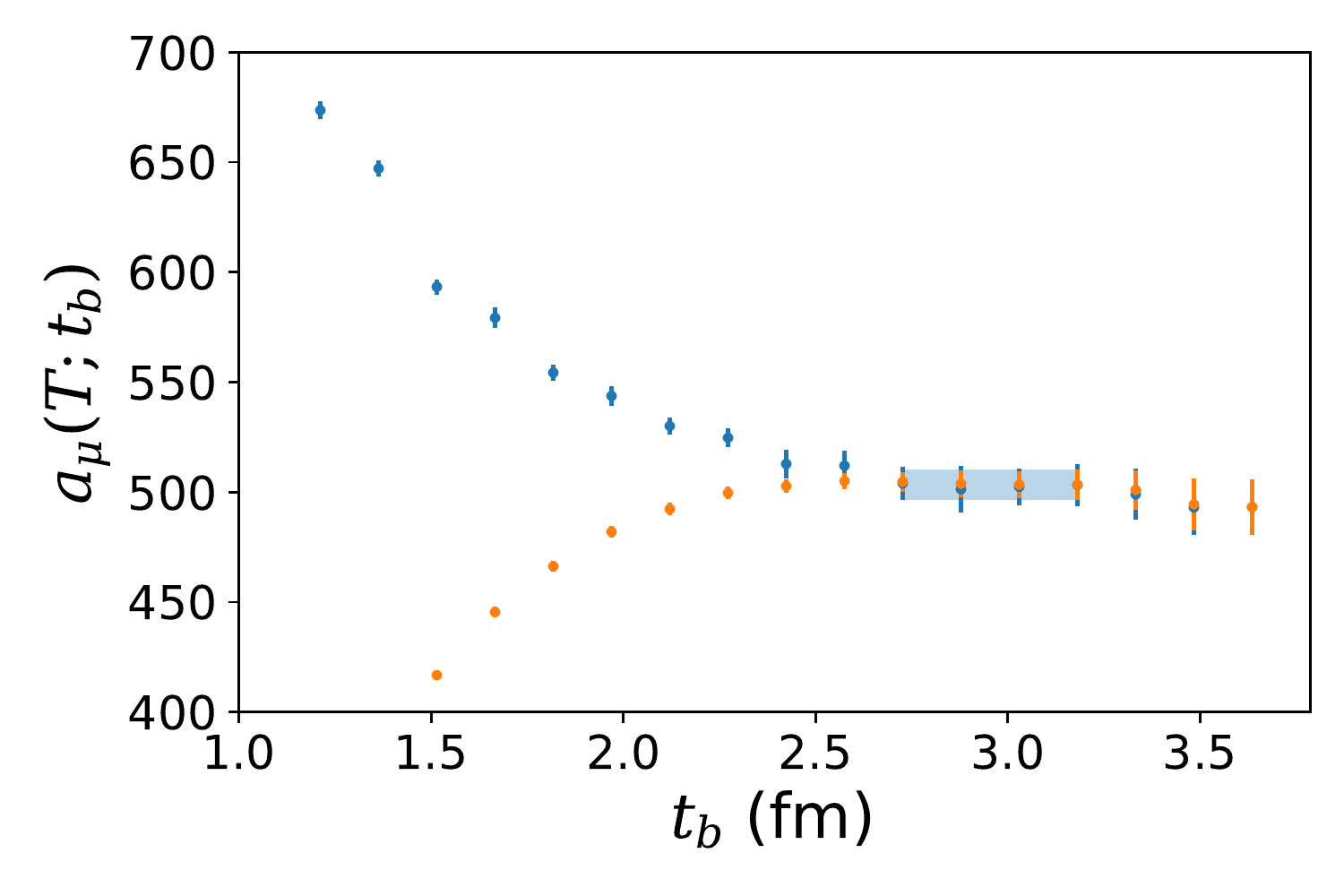}
\vspace{0.0cm}
\includegraphics*[width=7cm]{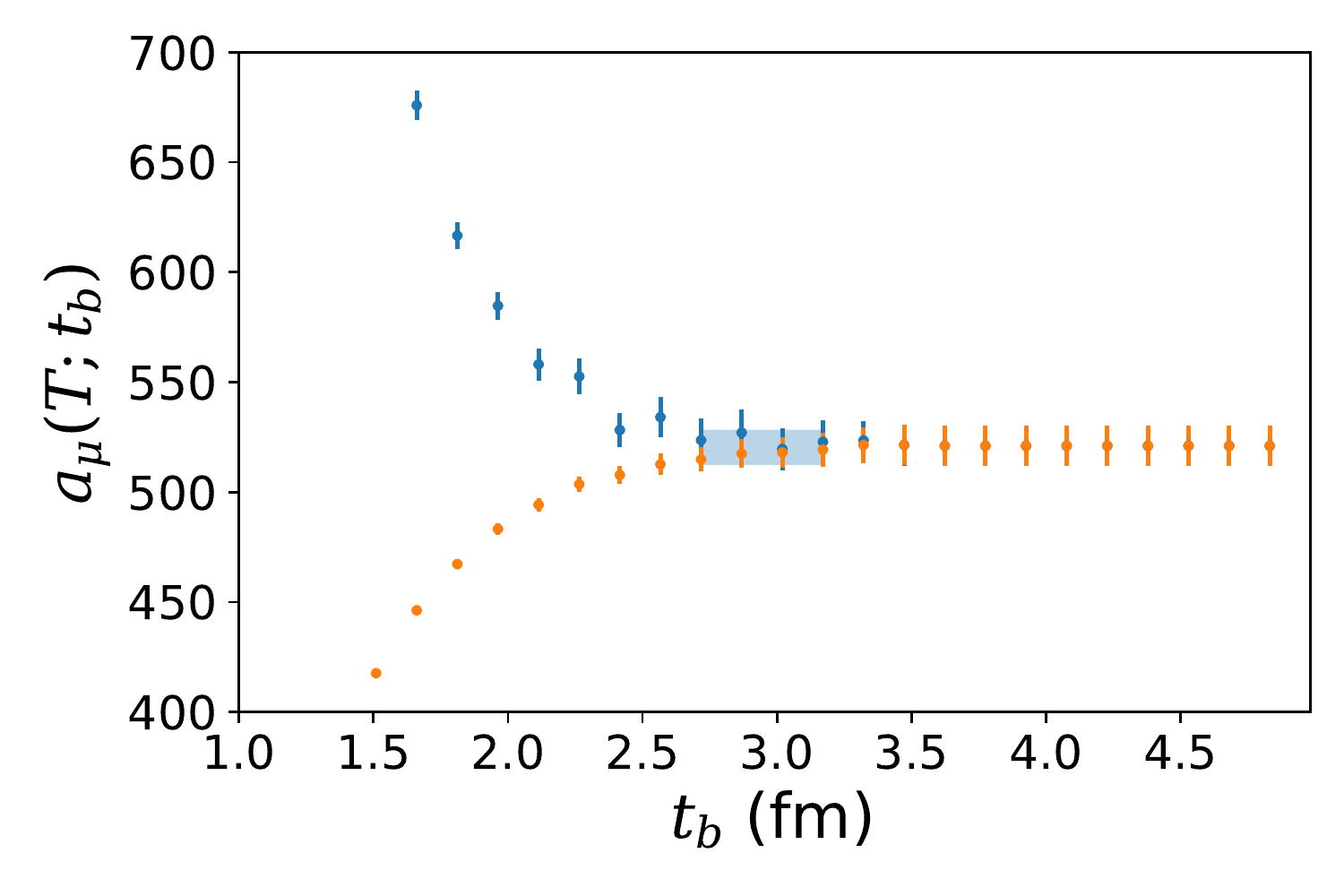}
\end{center}
\begin{quotation}
\floatcaption{fig:results}%
{{\it $a_\m(T;t_b)$ for the connected light quark contribution to the muon anomaly, using the bounding method of Ref.~\cite{RBC18}, as a function of the switch-point $t_b$, for the 96 (upper left), 64 (upper right), 48I (middle left), 32 (middle right), and 48II (lower) ensembles. 
Values obtained with the upper bound are shown in blue; values obtained with the lower bound are
shown in orange.  The values in Table~\ref{tab:results} are obtained from the shaded regions in each of these plots.  See
text.}}
\end{quotation}
\vspace*{-4ex}
\end{figure}

\begin{table}[t]
\begin{center}
\begin{tabular}{|c|c|c|c|} 
\hline
ensemble & $a_\mu^{\rm HVP,lqc}$ & $a_\m^{\rm W1,lqc}$  & $a_\m^{\rm W2,lqc}$  \\
\hline
96 & 606.1(13.0)(5.8) & 205.92(38)(45) & 94.7(2.4)(1.5) \\
64 & 596.5(9.0)(5.6) & 204.97(26)(49) & 88.1(1.3)(1.4) \\
48I & 547.0(8.6)(5.3) & 201.42(53)(56) & 76.8(1.4)(1.4) \\
32 & 503.3(6.9)(4.9) & 200.49(68)(60) & 70.8(1.0)(1.3) \\
48II & 520.4(8.0)(5.0) & 200.48(42)(60) & 71.4(1.2)(1.3) \\
\hline
\end{tabular}
\end{center}
\vspace*{-3ex}
\begin{quotation}
\floatcaption{tab:results}{{\it Results for $a_\mu^{\rm HVP,lqc}$, the $0.4-1.0$~fm window $a_\m^{\rm W1,lqc}$ and the $1.5-1.9$~fm window $a_\m^{\rm W2,lqc}$.   The first error is statistical, the second error from scale setting.}}
\end{quotation}
\vspace*{-4.5ex}
\end{table}

\begin{boldmath}
\section{\label{ChPT} Chiral perturbation theory for $a_\mu^{\rm HVP}$}
\end{boldmath}
In this section, we will present the NNLO result for $C(t)$ calculated in two-flavor SChPT, in a finite spatial volume $L^3$ with periodic boundary conditions (considering also finite-$T$ effects to NLO).\footnote{Nomenclature:  Because at leading non-trivial order in ChPT $C(t)$ 
is given by a one-loop diagram, we refer to this order as ``next-to-leading order''(NLO).} This extends our earlier calculation of Ref.~\cite{ABGP19} to include staggered effects also at NNLO.  Our results agree with those of Ref.~\cite{BMW20}.   
In Sec.~\ref{formulae} we present our result, in Sec.~\ref{comp} we compare NNLO ChPT with Ref.~\cite{Coletal} in order to estimate the maximal value of the pion mass at which NNLO ChPT can be trusted, and in Sec.~\ref{taste} we discuss taste splittings in more detail.

Before we embark on all this, we would like to emphasize again that ChPT can be
systematically applied to \amuHVP, to all orders \cite{ABGPEFT}, if one 
considers the EFT which extends ChPT for the strong interactions to include the electromagnetic coupling of pions to the external muon.   In this framework, counter terms beyond those present in the strong chiral lagrangian show up starting at NNNLO.   This can be seen by integrating over $t$ in Eq.~(\ref{amuHVP}), which leads to the representation \cite{LR,TB}
\begin{equation}
\label{amuHVPmom}
a_\m^{\rm HVP}=4\a^2\int_0^\infty dQ^2\,f(Q)\,\hat\Pi(Q^2)\ ,
\end{equation}
where $\hat\Pi(Q^2)$ is the subtracted scalar vacuum polarization.  Since, in ChPT, $\hat\Pi(Q^2)\sim (Q^2)^{k-1}$ at N$^k$LO (modulo logarithms) for large $Q^2$, and $f(Q)\sim m_\m^4/Q^6$ for large $Q$, we see that the ChPT result is finite for $k=1$ and $k=2$, but that a new counter term is needed at $k=3$.  This counter term has the form \cite{ABGPEFT}
\begin{equation}
\label{ct}
\frac{\a^2 m_\m^3}{(4\p f_\p)^4}\, {\bar\m}\s_{\a\b}F_{\a\b}\m\,\mbox{tr}(Q\S Q\S^\dagger)\ ,
\end{equation}
where $\m$ is the muon field, $\S$ is the non-linear pion field, $F_{\a\b}$ is the electromagnetic field-strength tensor, and $Q$ is the charge matrix.   Since we will consider $C(t)$ to NNLO only, this counter term will not be needed below.  

\vskip0.8cm
\subsection{\label{formulae} Formulae}
We have calculated $C(t)$ to NNLO in SChPT.  In momentum space, the corresponding calculation was done before in Ref.~\cite{BR} in continuum ChPT, while $C(t)$ was calculated in SChPT to NNLO in Ref.~\cite{BMW20}, with which our result agrees.

SChPT differs from standard continuum ChPT, because the symmetry group of lattice QCD with staggered fermions is smaller than the continuum symmetry group.  Hence, new operators appear in the chiral lagrangian multiplied by powers of the lattice spacing $a$.\footnote{In addition, low-energy constants which already appear in the continuum chiral lagrangian will also become dependent on $a$.}    We follow a power-counting scheme in which powers
of $p^2$ (with $p$ a typical momentum), the quark mass $m$, and the square of the lattice spacing $a^2$ are of the same order.  Operators of order $a^2$ have been classified in Refs.~\cite{LS,AB}, those appearing at orders $a^2p^2$,
$a^2m$ and $a^4$ have been classified in Ref.~\cite{SvdW}.

In an NNLO calculation, we work to $\co(p^6)$.   Tree-level contributions to $C(t)$ only appear at $\co(p^4)$ and $\co(p^6)$ and lead to contact terms proportional to $\d(t)$.
Since $w(t)\sim t^4$ at small $t$, we can ignore such contact terms, unless they would contain four derivatives.   The only contact term leading to such a contribution is proportional to the low-energy constant (LEC) $c_{56}$ \cite{BCE}, and this contact term does lead to an NNLO contribution to \amuHVP.  For a more detailed discussion of this contact term, we refer to App.~A of Ref.~\cite{ABGPEFT}. 
Its contribution cancels in differences like FV corrections and taste-breaking corrections, which are our focus in this paper.\footnote{In principle, a term of order $a^5$ can appear in the SChPT lagrangian \cite{BKL}.  However, it does not contribute to \amuHVP\ to NNLO, as explained in Ref.~\cite{BMW20}.}

Each staggered fermion leads to four degenerate fermions (``tastes'') in the continuum limit.   If we consider two-flavor QCD with staggered fermions, we thus obtain eight fermions in the continuum limit, four up quarks and four down quarks.   The problem that there are too many sea quarks is resolved by taking the fourth root of each staggered fermion determinant, which effectively reduces the total number of fermions in the continuum limit by a factor four.  In SChPT, this is handled as follows.   One introduces $N$ staggered fermions of each flavor, and the theory thus has 2 (for flavor) $\times$ 4 (for taste) $\times$ $N$ quarks in the continuum limit (which, in the isospin limit, are fully degenerate).  One develops SChPT by considering the EFT for the $(8N)^2-1$ pions in this theory, using the $SU(8N)_L\times SU(8N)_R$ symmetry, and spurions to introduce the symmetry breaking effects of the quark masses and lattice spacing.   Finally, in order to reflect the fourth roots at the QCD level, one sets $N=1/4$.  (For a review of the validity of this procedure, as well as the validity of taking the fourth root, we refer to the review in Ref.~\cite{MGroot}.)

Keeping $N$ general, the NNLO expression for $C(t)$ in SChPT is, for $t>0$
(thus avoiding contact terms)
\begin{eqnarray}
\label{Ctexpr}
C(t)&=&\frac{N^2}{3}\,\frac{1}{V}\sum_{\vec p}\sum_X\frac{{\vec p}^2}{E^2_X(p)}\,e^{-2E_X(p)t}\Biggl(1
-\frac{N}{f^2}\sum_YD_Y(0)-\frac{16\ell_6(\vec p^2+m_X^2)}{f^2}\Biggr)\nonumber\\
&&+\frac{N^3}{36f^2}\,\frac{1}{V^2}\sum_{XY}\sum_{\vec p\,\vec q}\frac{{\vec p}^2{\vec q}^2}{E^2_X(p)E^2_Y(q)}
\,\frac{E_X(p)e^{-2E_Y(q)t}-E_Y(q)e^{-2E_X(p)t}}{\vec p^2-\vec q^2+m_X^2-m_Y^2}\\
&=&\frac{N^2}{3}\,\frac{1}{V}\sum_{\vec p}\sum_X\frac{{\vec p}^2}{E^2_X(p)}\,e^{-2E_X(p)t}\Biggl(1
-\frac{N}{f^2}\sum_YD_Y(0)-\frac{16\ell_6(\vec p^2+m_X^2)}{f^2}\nonumber\\
&&\hspace{2cm}+\frac{N}{6f^2}\,\lim_{\eta\to 0} \mbox{Re}\ \frac{1}{V}\sum_{\vec q}\sum_Y\frac{{\vec q}^2}{E_Y(q)}
\,\frac{1}{\vec q^2-\vec p^2-i\eta+m_Y^2-m_X^2}\Biggr)\ .
\nonumber
\end{eqnarray}
The sum over momenta are sums over integer vectors $\vec{n}$, with $\vec{k}=2\p\vec{n}/L$, reflecting the finite
spatial volume $L^3$ with periodic boundary conditions.
Sums over $X$ and $Y$ represent sums over all sixteen taste pions, for each of the $N$ replicas.   The exact staggered symmetry group implies that there is some degeneracy, and, for instance,
\begin{eqnarray}
\label{sumX}
\sum_X\frac{1}{E^2_X(p)}\,e^{-2E_X(p)t}
&=&\frac{e^{-2E_5|t|}}{E_5^2}+\frac{3e^{-2E_{k5}|t|}}{E_{k5}^2}
+\frac{e^{-2E_{45}|t|}}{E_{45}^2}
+\frac{3e^{-2E_{jk}|t|}}{E_{jk}^2}\\
&&\hspace{1.5cm}+\frac{3e^{-2E_{k4}|t|}}{E_{k4}^2}+\frac{3e^{-2E_k|t|}}{E_k^2}+\frac{e^{-2E_4|t|}}{E_4^2}+\frac{e^{-2E_s|t|}}{E_s^2}\ ,
\nonumber
\end{eqnarray}
where
\begin{equation}
\label{EX}
E_X(p)=\sqrt{m_X^2+\vec{p}^2}\ ,
\end{equation}
with $m_X$ the mass of pion with taste $X$.  The different taste pions are labeled by irreducible representations of the staggered symmetry group, 
with $X\in\{5,k5,45,jk,k4,k,4,S\}$ with degeneracies $1$, $3$, $1$, $3$, $3$, $3$, $1$ and $1$, respectively \cite{MGmesons}.  The physical pion mass
is $m_\p=m_{X=5}$; this is the pion that becomes massless at non-zero lattice spacing if the bare quark mass is taken to zero.
The heaviest pion in a taste multiplet is $m_S=m_{X=S}$.
The only two LECs appearing in Eq.~(\ref{Ctexpr}) are the pion decay constant in the chiral limit, $f$, and the $\co(p^4)$ LEC $\ell_6$.  In our convention, the physical pion decay 
constant is $f_\p=130.4$~MeV.  Finally, $D_Y(0)$ represents a tadpole loop:
\begin{equation}
\label{DY}
D_Y(0)=\frac{1}{V}\sum_{\vec k}\int\frac{dk_4}{2\p}\,\frac{1}{k_4^2+E_Y^2(k)}=\frac{1}{V}\sum_{\vec k}\frac{1}{2E_Y(k)}\ .
\end{equation}
The expression in Eq.~(\ref{Ctexpr}) is valid when a current in the taste-singlet representation is used.  This is the case for our simulations, where we always use 
conserved currents.

As we did in Ref.~\cite{ABGP19}, we use Poisson resummation to split the expression for $C(t)$ into its infinite and finite volume parts.  The infinite-volume part diverges, and is renormalized by $\ell_6$, with
($\e=3-d$, with $d$ the number of spatial dimensions in dimensional regularization)
\begin{eqnarray}
\label{ell6}
\ell_6&=&\ell_6^r(\m)+\frac{N}{12\p^2}\left(\frac{1}{\e}+\half\log{(4\p)}-\half\g_E+\half\right)\\
&\equiv&-\frac{N}{24\p^2}\left(\bar{\ell}_6(m_\p,N)+\log\frac{m_\p^2}{\m^2}\right)
+\frac{N}{12\p^2}\left(\frac{1}{\e}+\half\log{(4\p)}-\half\g_E+\half\right)\ ,\nonumber
\end{eqnarray}
where the latter equation defines $\bar{\ell}_6$, which for $N=1/4$ has the value $16.0(9)$ \cite{BCT}.

The result~(\ref{Ctexpr}) turns out to be quite simple: the effect of SChPT, in comparison with continuum ChPT is that each pion loop gets averaged over the tastes in each staggered multiplet.  There are no other effects of the many operators that appear in SChPT beyond ChPT (for some more discussion of the contribution of classes of operators unique to SChPT, see Ref.~\cite{BMW20}). We will use Eq.~(\ref{Ctexpr}) setting $f=f_\p=130.4$~MeV, $N=1/4$ and $\bar{\ell}_6(m_\p,1/4)=16$.   Furthermore, we will use values for the taste masses $m_X$ 
measured by MILC on the first four ensembles of Table~\ref{tab:ensembles}.\footnote{We thank Doug Toussaint for providing us with the complete taste spectra on these ensembles.}  For the fifth ensemble in Table~\ref{tab:ensembles} we will use the same taste splittings as for the $32^3$ ensemble, as these two 
ensembles have been generated using the same lattice action at (nearly) the same lattice spacing, and we expect taste splittings to be (nearly) independent of the volume.   Since the physical 
pion masses on these two ensembles are not exactly the same, we use the relation
\begin{equation}
\label{tastesplit}
\D M_{\rm taste}^2\equiv m_X^2-m_\p^2=m_X^2-m_5^2=\D_X\ ,
\end{equation}
with $\D_X$ the taste splittings on the $32^3$ ensemble,
for the conversion.   We then substitute Eq.~(\ref{Ctexpr}) into Eq.~(\ref{amuHVP}),
multiplying by a factor $10/9$ to reflect the fact that we are considering only the light-quark connected part on the lattice \cite{DMJ,Mainz13}, so that
\begin{equation}
\label{amuHVPconn}
a_\m^{\rm HVP,lqc}=\frac{10}{9}\,2\int_0^\infty dt\ w(t) \,C(t)\ .
\end{equation}

Finally, we have considered finite-$T$ effects to NLO, finding that finite-$T$ effects are always much smaller than our statistical errors, and we will thus not take these into account.

\subsection{\label{comp} Range of validity}

\begin{figure}[t!]
\vspace*{4ex}
\begin{center}
\includegraphics*[width=9cm]{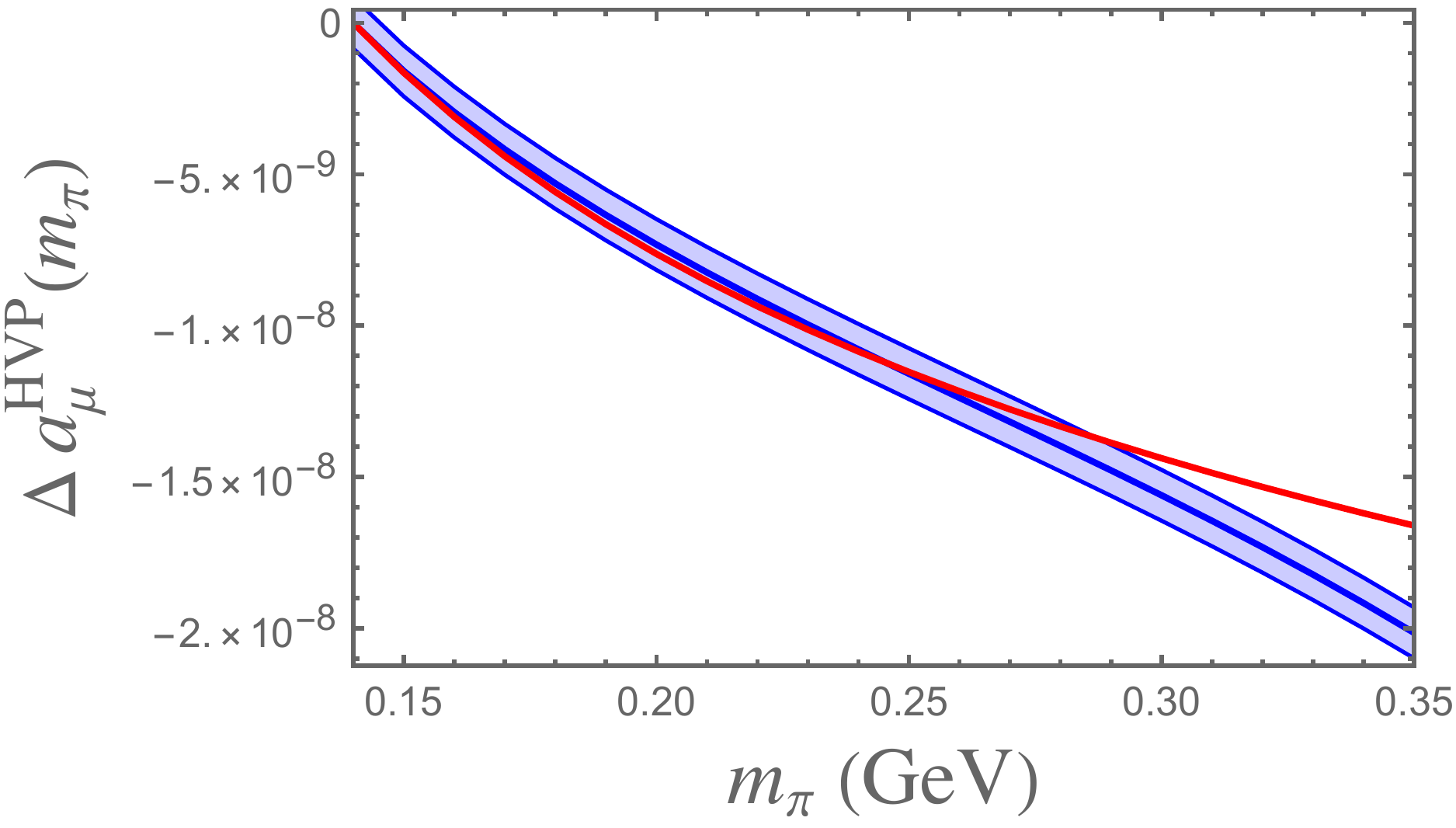}
\end{center}
\begin{quotation}
\floatcaption{compareNNLO}%
{{\it Comparison of $\D a_\m^{HVP}(m_\p)=a_\m^{HVP}(m_\p)-a_\m^{HVP}(m_\p=140\ \mbox{MeV})$ between NNLO ChPT, and the resummation of Ref.~\cite{Coletal}, as a function of $m_\p$.
The blue band gives the result of Ref.~\cite{Coletal}; the red curve is computed with NNLO ChPT.}}
\end{quotation}
\vspace*{-4ex}
\end{figure}

Let us begin with the ChPT view, to NNLO, of \amuHVP\ in the continuum limit, in infinite volume.   For this, it is easiest to use the momentum representation, Eq.~(\ref{amuHVPmom}).  The expression for the scalar vacuum
polarization to NNLO can be found in Refs.~\cite{GK,ABT}, and the value for
$c_{56}$ can be obtained from the $SU(3)$ LEC $C_{93}$ by matching \cite{SU3SU2}, with $C_{93}$ determined in Ref.~\cite{GMP}.  In the ``$\overline{\rm MS}+1$'' scheme used in all these references, 
\begin{equation}
\label{c56}
c_{56}(m_\r)=-1.3(4)\times 10^{-4}\ .
\end{equation}
Using Eq.~(\ref{amuHVPmom}), we find\footnote{Since this is computed in two-flavor ChPT, this reflects the two-pion contribution, plus the effects of the $\r$, through $\ell_6$.  The contribution from the pion form factor is believed to be about $5\times 10^{-8}$, and it is not clear how much other physics (such as that from kaon physics, for example) is represented through the values of the LECs $f_\p$, $\ell_6$ and $c_{56}$.   We will not elaborate on this point further here, as the error in Eq.~(\ref{amuHVP2flChPT}) is large.}
\begin{equation}
\label{amuHVP2flChPT}
a_\m^{\mathrm{HVP,2-flavor\ ChPT}}=6.6(1.6)\times 10^{-8}\ .
\end{equation}
The central value is very reasonable, but the error, mostly due to the error in $c_{56}$, is very large.   To NLO in ChPT, we find instead of Eq.~(\ref{amuHVP2flChPT}) the value $0.7\times 10^{-8}$.  This indicates that NNLO ChPT may give a good understanding of \amuHVP, even if, as is well known, NLO ChPT does not. The reason is that the $\r$ meson only starts contributing to the pion form factor beyond leading order.  This also suggests that the apparent lack of convergence is not necessarily a reason to worry about the applicability of ChPT.

We can do much better.   First, if we are interested in corrections due to finite volume, pion mass mistuning or taste breaking, the LEC $c_{56}$ drops out, removing a significant source of error.  Second, in a recent paper a good representation of the two-pion contribution to \amuHVP\ was obtained using a resummation of NNLO ChPT based on unitarity and analyticity \cite{Coletal}.   In Ref.~\cite{Coletal} this was used to study the dependence 
of \amuHVP\ on the pion mass.   We can do the same using straight
NNLO ChPT, and compare to Ref.~\cite{Coletal}.

The comparison is shown in Fig.~\ref{compareNNLO}, which shows the difference between \amuHVP\ values at different pion masses.  The band in this figure represents the estimate obtained in Ref.~\cite{Coletal},\footnote{We thank Martin Hoferichter for providing us with the code to reproduce this plot.} while the curve represents NNLO ChPT.  We see that if we take the result from Ref.~\cite{Coletal} as a benchmark, NNLO ChPT does remarkably well up to pion masses of about $250$~MeV.  This suggests that NNLO ChPT can be used safely to compute pion mass retunings,
as well as corrections for taste splittings, as long as the taste splittings are not too large.  From Table~\ref{tab:ensembles}, we see that all pion masses in the taste multiplet are well within this range for the 96 and 64 
ensembles.   While the 48I ensemble would not appear to 
qualify, one should bear in mind that $m_S=326$~MeV is the mass of the heaviest member of the taste multiplet, which counts for only 1/16-th of the average in Eq.~(\ref{sumX}).  The root-mean-square mass on this ensemble is 
equal to $241$~MeV, and it is thus possible that NNLO ChPT is reliable for 
this ensemble as well.  

Of course, also the result from Ref.~\cite{Coletal} is based on NNLO ChPT, 
using a resummation based on the Omn\`es relation.   However, Ref.~\cite{Coletal}
carried out a careful comparison of the prediction from resummed NNLO ChPT
with the physical value of the two-pion contribution to \amuHVP\ at the physical pion mass,\footnote{Taking the physical pion mass to be 140~MeV,
which is why we take the reference mass in Fig.~\ref{compareNNLO} to be
140~MeV.} with excellent agreement.   

\begin{figure}[t!]
\vspace*{4ex}
\begin{center}
\includegraphics*[width=7cm]{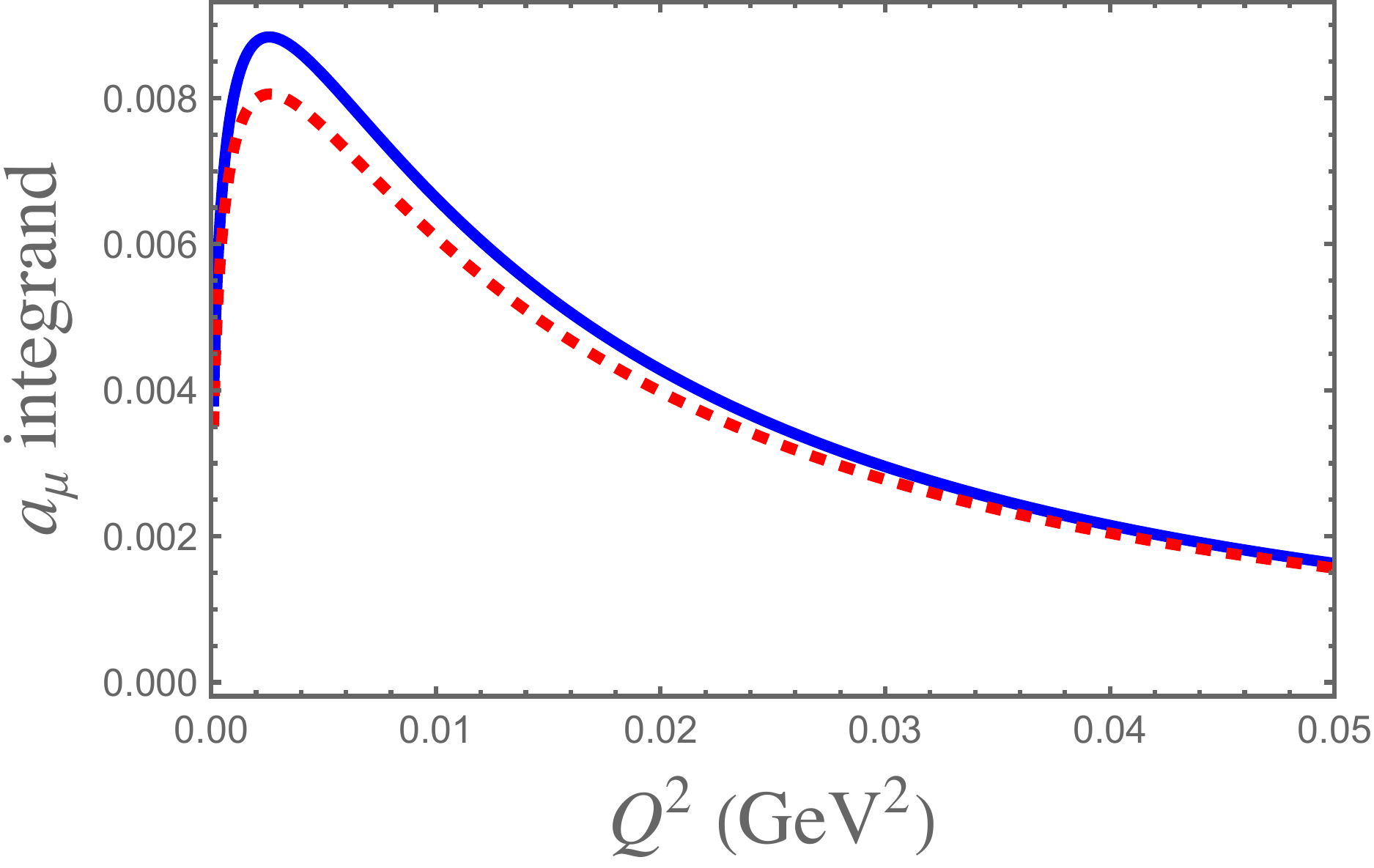}
\hspace{0.5cm}
\includegraphics*[width=7cm]{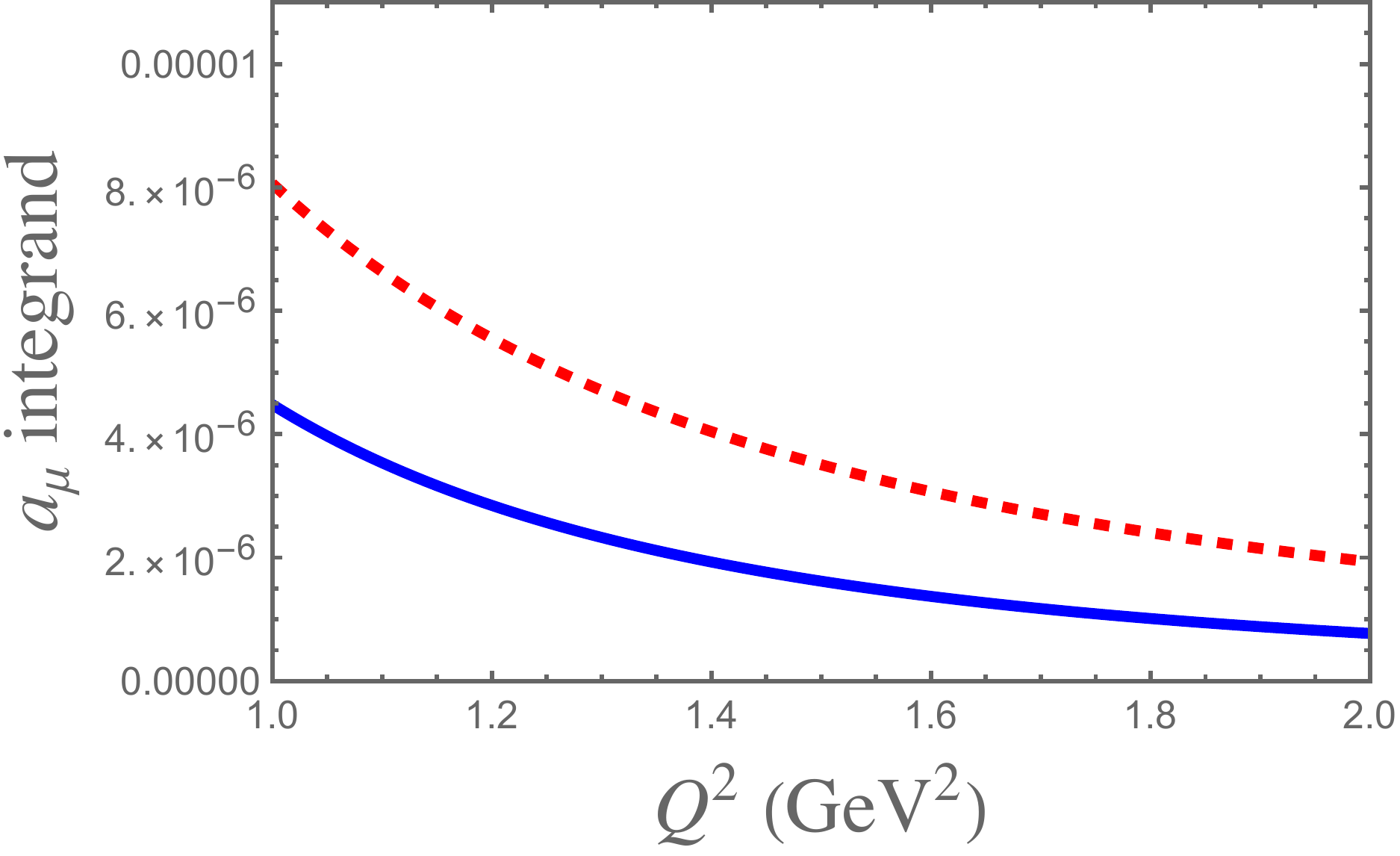}
\end{center}
\begin{quotation}
\floatcaption{intamu}%
{{\it The integrand for Eq.~(\ref{amuHVPmom}).  The solid blue curve is obtained from the R ratio compiled in Ref.~\cite{KNT18}, the dashed red curve is obtained from NNLO ChPT, with parameters as specified in the text.  The panel on the left shows the region up to $Q^2=0.05$~{\rm GeV}$^2$ and the panel on the right shows the region $1$ to $2$ {\rm GeV}$^2$.}}
\end{quotation}
\vspace*{-4ex}
\end{figure}

The situation is, in principle, different for window quantities.\footnote{As also observed in Ref.~\cite{BMW20}.}  Qualitatively, this can be understood from
Fig.~\ref{intamu}, which shows the integrand of Eq.~(\ref{amuHVPmom}).   This figure explains why NNLO ChPT provides a decent approximation to the full integral from $Q^2=0$ to $\infty$: the ChPT curve is close to the curve derived from data.  However, cutting out a window in time, with the window function~(\ref{windowdef}), corresponds to cutting out a region in $Q^2$, with a short-distance window corresponding to a large-$Q^2$ region.  If the $Q^2$ values in a region emphasized by a certain window are large enough, the right-hand panel in Fig.~\ref{intamu} shows why NNLO ChPT may yield a bad prediction for the window quantity. 
For example, the NNLO ChPT value for a window emphasizing $Q^2$ values between $1$ and $2$ GeV$^2$ might overshoot the real value by a factor of about two,
from the right-hand panel in Fig.~\ref{intamu}.  NNLO ChPT is
expected to work better for a window with larger values of $t_0$ and $t_1$.

\subsection{\label{taste} Taste splittings}
It is interesting to consider the behavior of taste splittings in the pion spectrum, as a function of the lattice spacing.  In Fig.~\ref{tastespectrum}, we
show the taste splittings $\D_X$ in Eq.~(\ref{tastesplit}) as a function of $a^2\a_s^2(1/a)$ for the first four ensembles in Table~\ref{tab:ensembles}, where $\a_s(1/a)$ is the $\overline{\mbox{MS}}$ coupling at the scale $1/a$.   The curves represent fits of the
form
\begin{equation}
\label{tastefit}
\D_X=A_X\a_s^2(1/a)a^2+B_Xa^4+C_Xa^6\ ,
\end{equation}
with different coefficients for each of the tastes $X\in\{k5,45,jk,k4,k,4,S\}$.\footnote{The scale in $\a_s$ is expected to be of order $1/a$.  Our fits are not sufficiently sensitive to the coefficient of $1/a$ inside $\a_s$ to fix this coefficient.} As we do not have access to the correlations between the different taste splittings on any given ensemble, our fits do not take correlations into account, and thus we will have to judge the fits visually.

At leading order, SChPT for HISQ fermions predicts the behavior reflected by the first term, with coefficient $A_X$ in Eq.~(\ref{tastefit}) \cite{MILC}. It also predicts that, at that order, the various taste multiplets fall into 
representations of $SO(4)$, \ie, that the masses with tastes $k5$ and $45$, with tastes $jk$ and $k4$ and with tastes $k$ and $4$ become pairwise degenerate \cite{LS}.
The other two terms, with coefficients $B_X$ and $C_X$ should be considered as phenomenological.   

From Fig.~\ref{tastespectrum}, several observations can be made.   First, indeed, the approximate $SO(4)$ multiplets at order $a^2$ are clearly seen.
However, in carrying out the fits, we find that the $A_X$ coefficients are not distinguishable from zero within errors. This is puzzling, because if the HISQ action
suppresses $\co(a^2)$ taste breaking to the extent that taste splittings do not follow an $\sim{a^2}$ behavior, one would also not expect to see the approximate
$SO(4)$ degeneracies.
Most important, it is clear that both $a^4$ and $a^6$ terms are needed in the fit, thus showing that, for these four ensembles, the behavior of the taste splittings is highly non-linear in $a^2\a_s^2(1/a)$.   Moreover, the blow-up on the right in Fig.~\ref{tastespectrum} shows that this non-linearity persists for the 96 and 64 ensembles, which have the smallest lattice spacings.   Indeed, we find that $a^4$ terms are still needed even if one tries to fit the taste splittings on the 96 and 64 ensembles only.
In a fit to only these two ensembles, we find that the $A_X$ coefficients are marginally different from zero.\footnote{For these fits, we set $A_{k5}=A_{45}$, $A_{jk}=A_{k4}$ and $A_k=A_4$ to have a positive number of degrees of freedom, consistent with leading-order SChPT.}  The main lesson from this discussion is that the behavior of taste splittings on these four ensembles is very far from 
linear in $a^2\a_s^2(1/a)$.  This is visually clear from Fig.~\ref{tastespectrum}.\footnote{It might be interesting to see whether these taste splittings can be better understood using NLO SChPT, for which the necessary calculations have been carried out in Ref.~\cite{BKL}.  This is beyond the scope of the present paper.}

\begin{figure}[t!]
\vspace*{4ex}
\begin{center}
\includegraphics*[width=7cm]{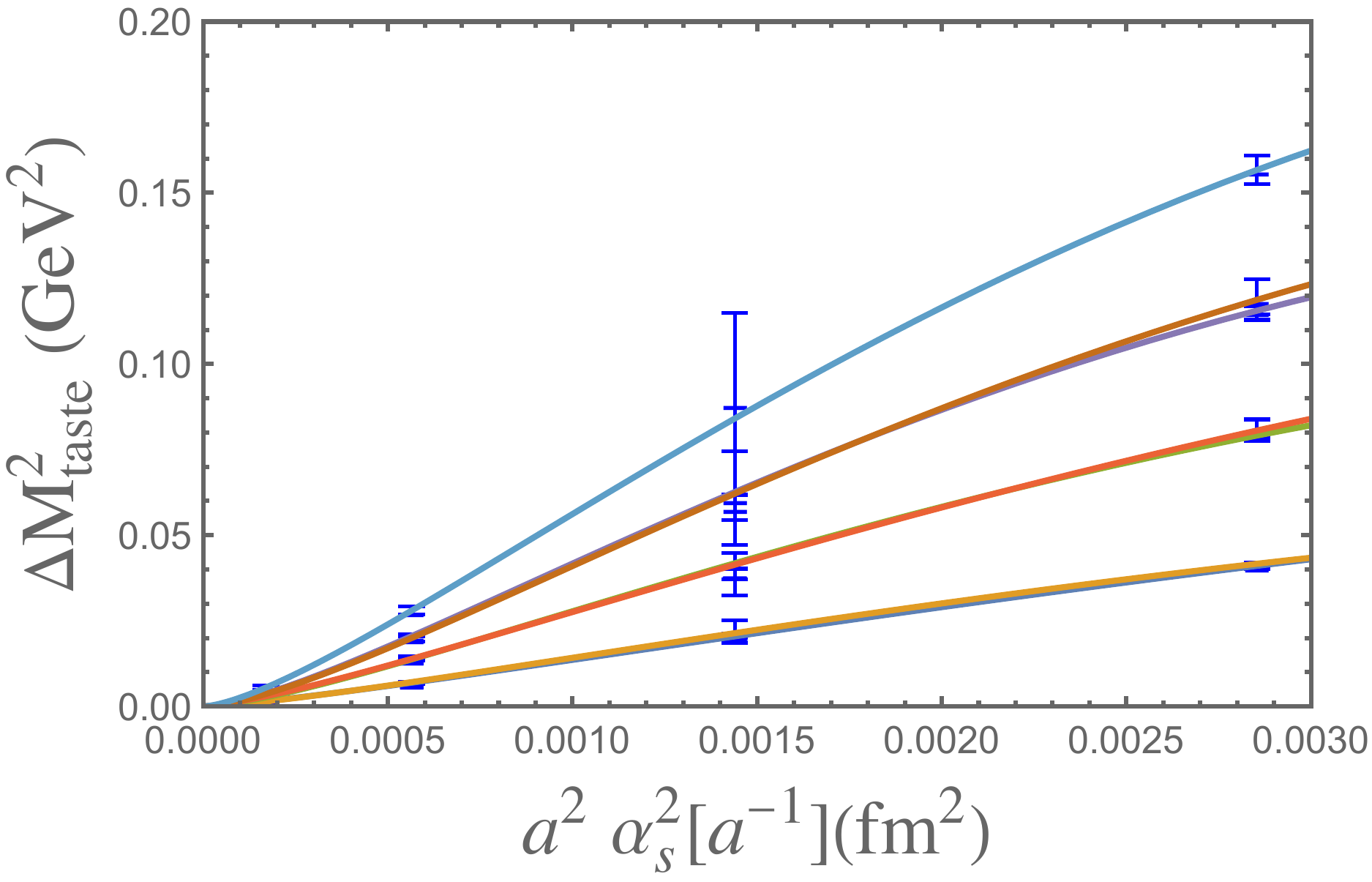}
\hspace{0.5cm}
\includegraphics*[width=7cm]{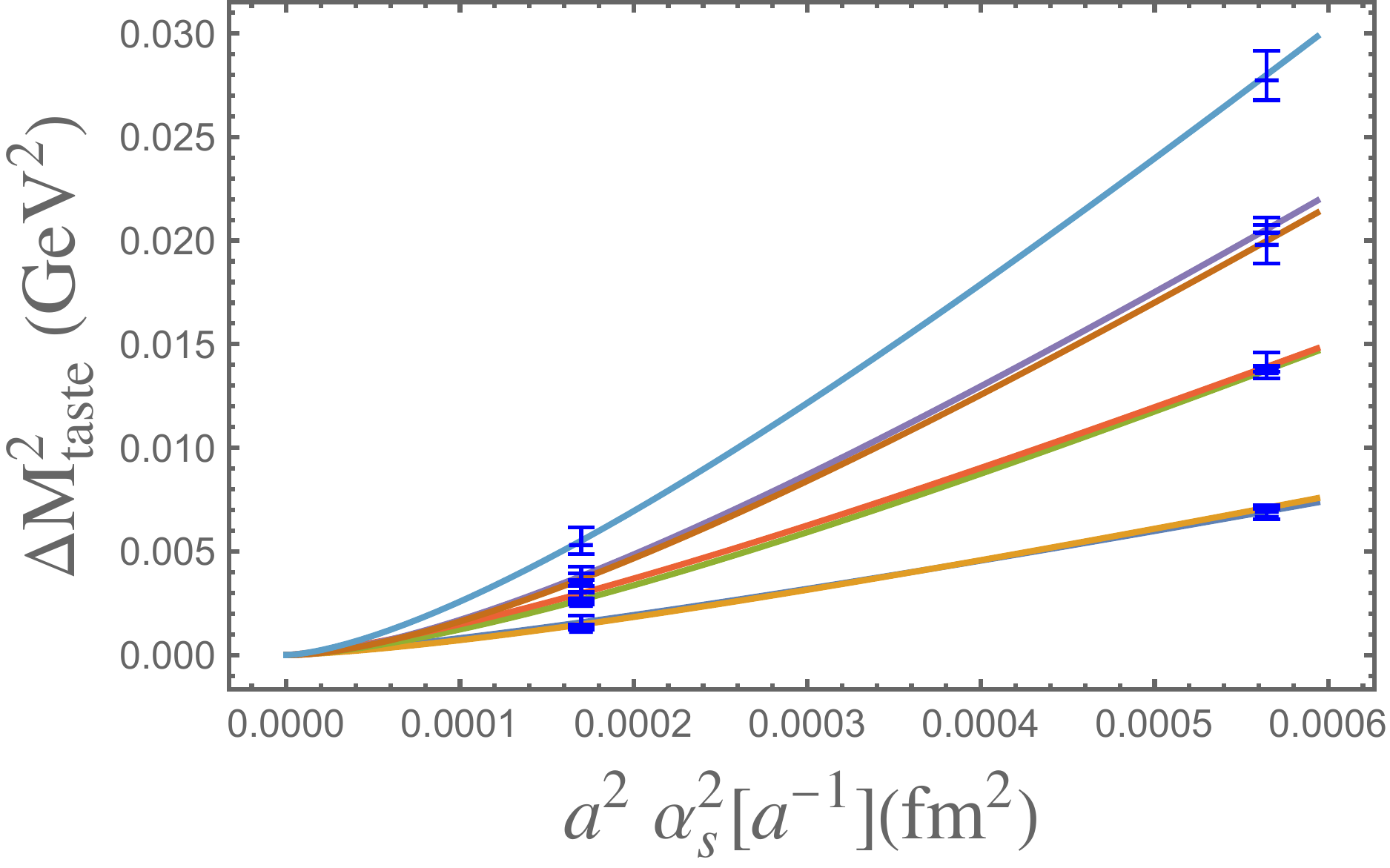}
\end{center}
\begin{quotation}
\floatcaption{tastespectrum}%
{{\it Pion taste splittings as a function of $a^2\a_s^2(1/a)$, with
$\a_s(1/a)$ the $\overline{{\rm MS}}$ coupling at scale $1/a$.  Points with error bars correspond to the measured taste splittings on the first four ensembles in Table~\ref{tab:ensembles}. The curves reflect fits to the form given in Eq.~(\ref{tastefit}); see main text.  The figure on the right zooms in on the region with the two smaller lattice spacings in Table~\ref{tab:ensembles}.  The near-degeneracy of the $X=k$ and $X=4$, $X=jk$ and $X=k4$, and $X=k5$ and $X=45$ taste
splittings is clearly visible (see text).}}
\end{quotation}
\vspace*{-4ex}
\end{figure}

\begin{boldmath}
\section{\label{full} Light-quark connected contribution to $a_\mu^{\rm HVP}$}
\end{boldmath}

\begin{table}[t]
\vspace*{4ex}
\centering
\begin{tabular}{|c|c|c|c|c|c|}
\hline
ensemble & 96 &  64  & 48I & 32 & 48II  \\
\hline
FV NLO   & 15.6 & 7.0 &  2.11 & 2.24 & 0.588\\
\hline
FV NNLO & 22.4(3.0) & 10.2(1.5) &  3.03(40) & 3.36(56)  & 0.751(45)\\
\hline\hline
FV SRHO model & 13.2 & 5.9 & 1.81 & 1.94 & 0.517 \\
\hline
\hline
\hline
retuning NLO & -0.80 & -6.24 &  -2.55 & -2.19 & -0.75\\
retuning NNLO & -1.38(42) & -10.6(3.1)   & -4.4(1.3)  & -3.8(1.1) & -1.29(39) \\
\hline
\hline
retuning NNLO no $\rho$ & -0.89 & -6.9  & -2.8 & -2.4 & -0.84\\
\hline
\hline
retuning SRHO model & -1.68 & -12.93 & -5.34 & -4.59 & -1.57 \\
\hline
\end{tabular}
\vskip3ex
\floatcaption{tab:FVmis}{{\it FV and pion-mass retuning corrections for $a_\mu^{\rm HVP,lqc}$, in units of $10^{-10}$, computed in SChPT and in the SRHO model. 
For the ``no $\rho$'' column, the value of $\bar{\ell}_6$ is reduced
from $16$ to $3.5$.  To be added to correct lattice results; for details, see text.}}
\end{table}

We begin this section by considering the various corrections to the lattice results provided by NNLO SChPT.  The FV and pion-mass-mistuning corrections are
shown in Table~\ref{tab:FVmis}; we also show the SRHO-model values for these corrections.  The table shows the NLO and NNLO FV and
pion-mass retuning corrections. All ``NNLO'' results will always be understood to include the NLO contribution as well. We estimated the errors from the truncation of ChPT geometrically from the NLO and NNLO values, and we show these errors with the NNLO results.  We expect the FV effects to be dominated by two-pion states, and indeed, the table shows a reasonably good convergence for all five ensembles.   For the retuning corrections, the convergence is somewhat less good.  In this case, we also show NNLO results one would obtain by replacing the
value 16 for $\bar{\ell}_6$ with 3.5.   This is an estimate for the value of $\bar{\ell}_6$ obtained by leaving out the effect of $\r$ by 
setting $\ell^r_6(\m=m_\r)=0$ in Eq.~(\ref{ell6}),
and thus, by comparison, gives an idea about the size of the $\r$ contribution, which enters only at NNLO through $\bar{\ell}_6$ \cite{GL1984}.  The ``retune NNLO no $\r$'' line in the table shows that without the $\r$ NNLO results are much closer to NLO results, thus confirming that the apparent poor convergence originates from the contribution of the $\r$.   The errors shown with the NNLO results are thus most likely conservative.   

\begin{table}[t]
\vspace*{4ex}
\centering
\begin{tabular}{|c|c|c|c|c|c|}
\hline
ensemble & 96 & 64 &  48I & 32 & 48II \\
\hline
t. br. NLO & 9.4  & 34.6  &  52.2 & 62.5 & 61.3 \\
t. br. NNLO & 16.6 & 65.8 &  114.0 & 151.6 & 149.6 \\
\hline
\hline
t. br. NNLO no $\rho$   & 10.5 & 39.2 &  60.2 & 72.9 & 71.4 \\
\hline
\hline
t. br. SRHO model & 18.1 & 71.4 & 122.5 &161.3 & 159.0 \\
\hline
\end{tabular}
\vskip3ex
\floatcaption{tab:tb}{{\it Taste breaking corrections in $a_\mu^{\rm HVP,lqc}$ in infinite volume, units of $10^{-10}$, computed in SChPT and in the SRHO model.
For the ``no $\rho$'' column, the value of $\bar{\ell}_6$ is reduced
from $16$ to $3.5$.  To be added to correct lattice results; see text.}}
\end{table}

Taste-breaking effects (in infinite volume) are shown in Table~\ref{tab:tb}.  In this case, SChPT does not appear to converge, although the
``no $\r$'' line suggests that the appearance of the $\r$ at NNLO is the likely reason for this lack of convergence.  However, for the coarser
ensembles, the largest taste masses are quite large, and this may be an additional reason for the poorer convergence for these ensembles.   
Below, we will consider continuum extrapolations with and without taste-breaking corrections.  As they are a lattice artifact, their effect 
should extrapolate away in the continuum limit.\footnote{This is also the reason we make no attempt to estimate ChPT truncation errors for the
taste-breaking effects.}

\begin{figure}[t!]
\vspace*{4ex}
\begin{center}
\includegraphics*[width=7cm]{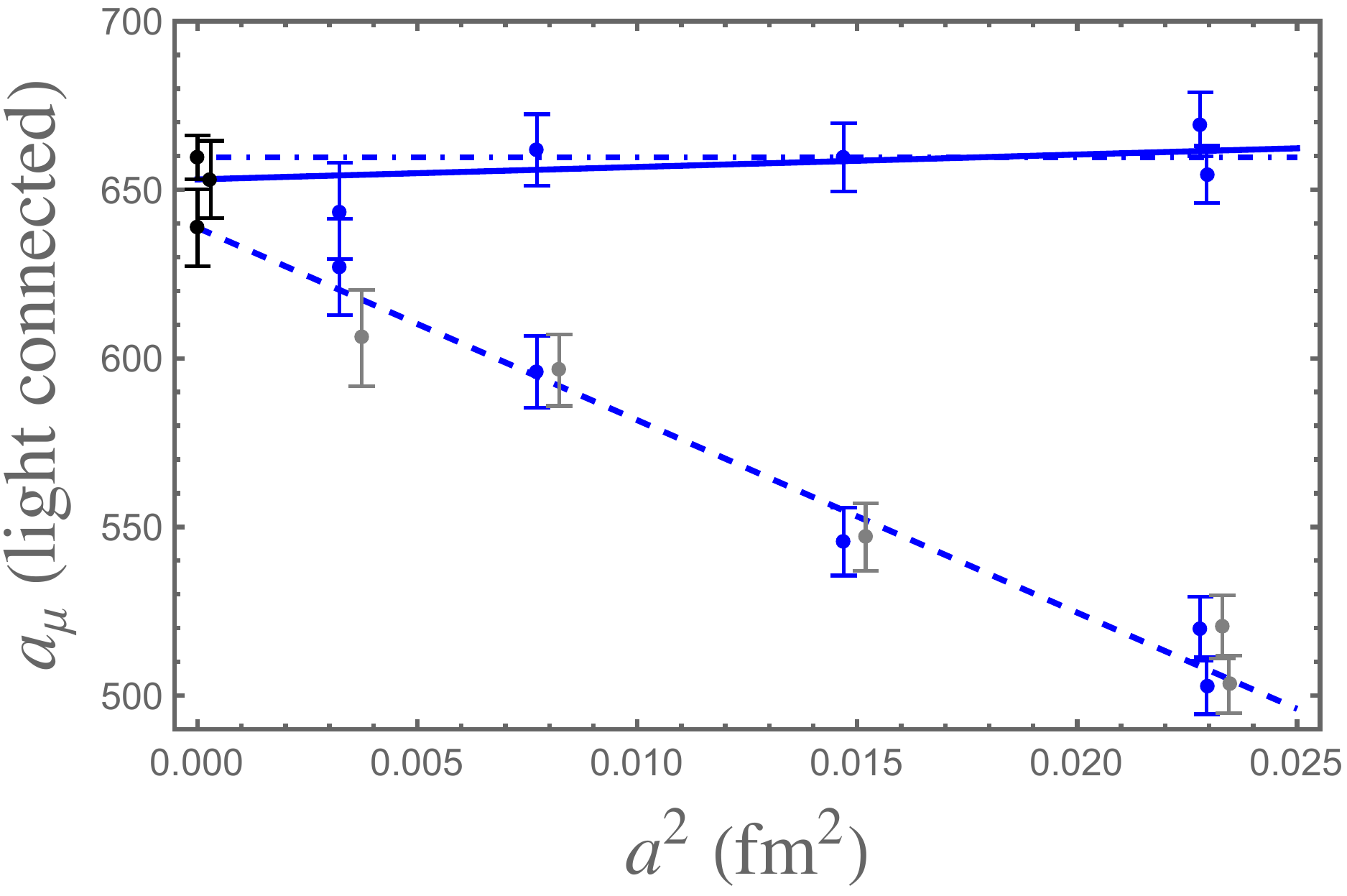}
\hspace{0.5cm}
\includegraphics*[width=7cm]{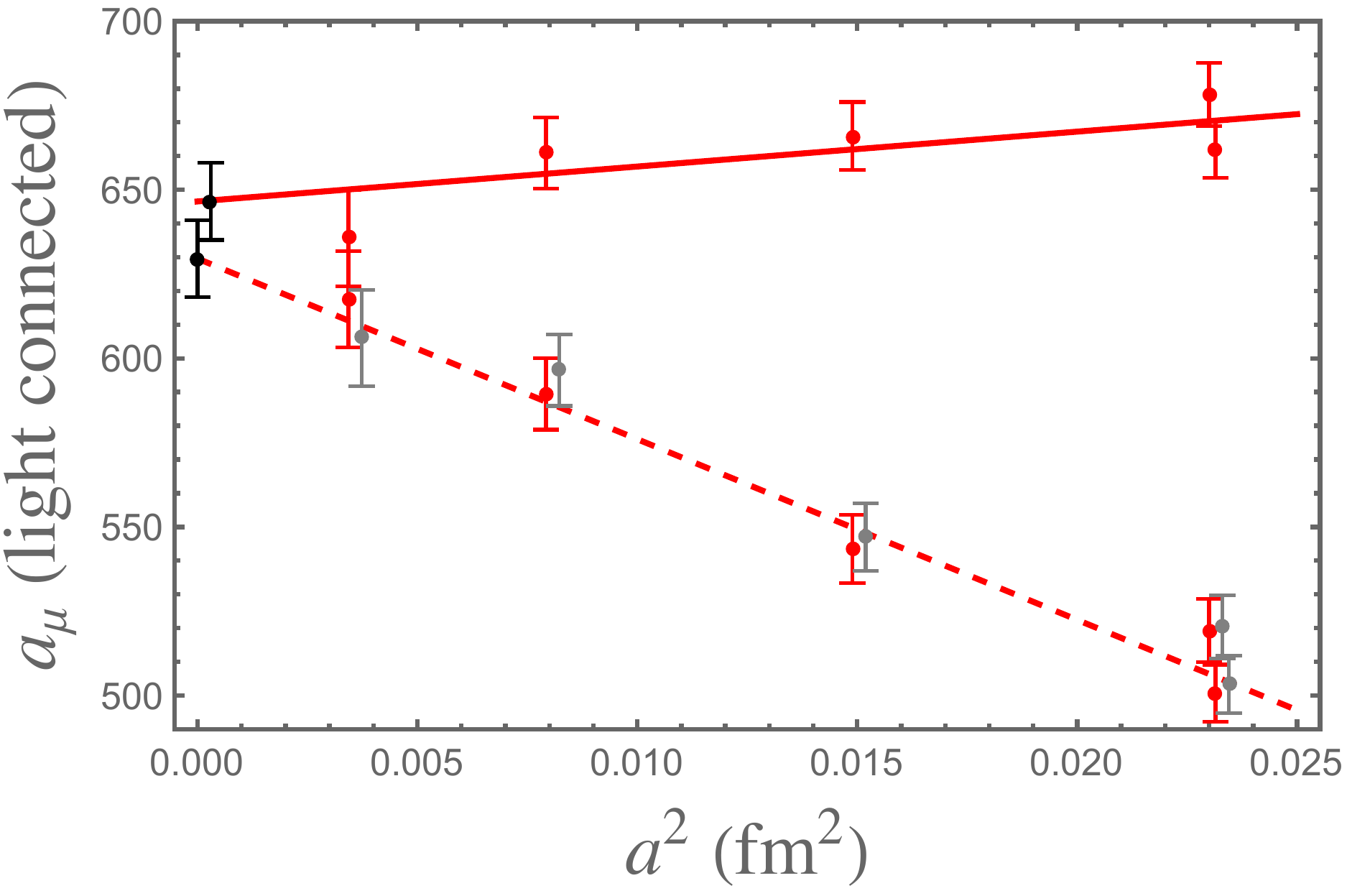}
\end{center}
\begin{quotation}
\floatcaption{amu}%
{{\it The continuum limit of $a_\mu^{\rm HVP,lqc}$ in units of $10^{-10}$. Left panel:  Fits using NNLO SChPT for FV corrections and pion mass retuning; without taste breaking  (linear, blue dashed line) and with taste breaking (linear, blue solid line or constant, blue dot-dashed line).   Right panel:
Fits using the SRHO model for FV corrections and pion mass retuning; without taste breaking  (linear, red dashed line) and with taste breaking (linear, red solid curve).   The data points for each fit are shown in the same color; continuum limits are shown in black.   The grey points in both panels (slightly horizontally offset for clarity) show the uncorrected values reported in Table~\ref{tab:results}.
Some continuum-limit
extrapolations are slightly offset for clarity.}}
\end{quotation}
\vspace*{-4ex}
\end{figure}

In Fig.~\ref{amu} we show our results for $a_\mu^{\rm HVP,lqc}$, using the values of Tables~\ref{tab:results}, \ref{tab:FVmis} and \ref{tab:tb}.  The gray data points show the
values of Table~\ref{tab:results}.  The FV
and pion-mass mistunings of Table~\ref{tab:FVmis} have to be applied before the continuum limit can be taken, of course.  The nearby colored points show the same data, corrected for FV effects and pion-mass mistuning using NNLO SChPT (left panel) or the SRHO model (right panel), but not for taste breaking.   Since taste breaking is an
order-$a^2$ effect, these corrected data allow us to extrapolate to the continuum limit, which is done with a linear fit in each panel, shown as a dashed line.     The 32- and 48II-ensemble points, which have different volumes but the same lattice spacing, agree within errors.  

In the left panel, we also show the data points corrected for taste breaking in NNLO SChPT; the solid line shows a linear fit, while the dot-dashed line shows a constant fit; both fits are in good agreement with the data.   In the right panel
data points also corrected for taste breaking are shown using the SRHO model; the solid line shows a linear fit to these SRHO-corrected data points.   We fit as a function of $a^2$, and not $a^2\alpha_s^2$, as the use of the trapezoidal integration in Eq.~(\ref{amut}) introduces an $a^2$ error.

For all fits, we have assumed the statistical errors to be uncorrelated, since the data are obtained on different ensembles.   The scale-setting errors have been assumed to be 100\% correlated (but assuming the latter to be uncorrelated does not have a big effect on the fits).  $P$-values for all fits are larger than 22\% (with the SRHO-based fit including taste-breaking corrections at this value).
All fits are good, and the data do not allow us to distinguish.
While the constant fit to the SChPT-corrected data gives the smallest error,
$\pm 6.4\times 10^{-10}$, clearly one cannot conclude that the constant fit
is preferred.  

Our extrapolated values from these fits are as follows:
\begin{eqnarray}
\label{amufits}
a_\m^{\rm HVP,lqc}&=& 638.7(11.4)\times 10^{-10}\ ,\qquad \mbox{NNLO~without~taste~breaking}\ ,\\
&=&653.1(11.4)\times 10^{-10}\ ,\qquad \mbox{NNLO}\ ,\nonumber\\
&=&659.6(6.4)\times 10^{-10}\ ,\qquad\ \,\mbox{NNLO~(constant)}\ ,\nonumber\\
&=&629.6(11.4)\times 10^{-10}\ ,\qquad \mbox{SRHO~without~taste~breaking}\ ,\nonumber\\
&=&646.5(11.4)\times 10^{-10}\ ,\qquad \mbox{SRHO}\ .\nonumber
\end{eqnarray}
We have also carried out linear fits dropping the two data points at $a=0.15$~fm, \ie, dropping the 32 and 48II ensembles.  The continuum-extrapolated
values differ from the central values above by not more than the fit errors in Eq.~(\ref{amufits}).  We note, however,
that the values with and without taste-breaking corrections become much closer.

As our best value, we take the average of the first two fits in Eq.~(\ref{amufits}), adding a systematic error equal to half the distance.   This
procedure is motivated by the fact that the two fits should agree in the continuum limit, since they differ only in the treatment of taste-breaking
effects.
Our best value for $a_\m^{\rm HVP,lqc}$
is then
\begin{equation}
\label{amuresult}
a_\m^{\rm HVP,lqc}=(645.9\pm 11.4\pm 7.2\pm 3.0\pm 0.4)\times 10^{-10}=646(14)\times 10^{-10}\ .
\end{equation}
The four errors are the statistical error from the fit, half the distance between the first two fits in Eq.~(\ref{amufits}), and the errors on the NNLO FV and
retuning corrections for the 96 ensemble shown in Table~\ref{tab:FVmis}, taking for the latter two error estimates those for the ensemble closest
to the continuum limit.   The error in the second equality in Eq.~(\ref{amuresult}) is obtained by adding these four errors in quadrature.  Applying the same procedure to the two SRHO-based values in Eq.~(\ref{amufits}) 
yields $638(14)\times 10^{-10}$, which is consistent with Eq.~(\ref{amuresult}).\footnote{No systematic errors associated with FV or retuning corrections can be
obtained in this case.}

Comparing this value with the value we obtained in Ref.~\cite{ABGP19}, $a_\m^{\rm HVP,lqc}=659(22)$, we make the following observations.  Our new central value is lower,
but consistent within errors.  Our total error has been reduced by a factor 1.6.  The main reason for this reduction is the reduction in statistical errors,
as can be seen by comparing Table~\ref{tab:results} with Table~II of Ref.~\cite{ABGP19}, especially for the $96^3$ ensemble.  Our scale setting error is now folded into
the fit error, as we took the scale-setting errors of Table~\ref{tab:results} into account in our fits, assuming them to be 100\% correlated.  
While the error in Eq.~(\ref{amuresult}) is 2.2\%, and thus still large relative to the sub-percent goal, we note that its central value is lower than the
corresponding value obtained in Ref.~\cite{BMW20}.\footnote{We estimate the value of Ref.~\cite{BMW20} by taking their value in finite volume,
$633.7(2.1)(4.2)\times 10^{-10}$, and adding $10/9$ (for the light-quark connected part) times their finite-volume correction $18.7(2.5)\times 10^{-10}$
(which we ascribe to the light-quark part), which yields $654.5(5.5)\times 10^{-10}$.}

There is good reason to believe that with these ensembles scaling violations are not linear in $a^2$, because already the taste splittings themselves are not linear in $a^2$,
as we showed in Sec.~\ref{taste}.  A linear extrapolation of the two smallest-$a$ central values of $a_\mu^{\rm HVP,lqc}$ leads
to a lower value of $a_\mu^{\rm HVP,lqc}$ in the continuum limit, $631\times 10^{-10}$, at the very low end of the range
in Eq.~(\ref{amuresult}).  Clearly, all these values are consistent with each other because of the relatively large fit errors. 
Because of this, it is interesting to look at window quantities.

\begin{table}[t]
\begin{center}
\begin{tabular}{|c|c|c|c|c|}
\hline
 & $a_\mu(96)-a_\mu(64)$ & $a_\mu(96)-a_\mu(\mbox{48I})$ & $a_\mu(96)-a_\mu(32)$ & $a_\mu(96)-a_\mu(\mbox{48II})$\\
\hline
lattice & 10(16) & 59(16) & 103(15) & 86(15)\\
\hline
NLO~SChPT & 11 & 28 & 38 & 37\\
NNLO~SChPT & 28 & 75 & 114 & 111 \\
\hline
SRHO & 35 & 89 & 129 & 128 \\
\hline
\end{tabular}
\end{center}
\vspace*{-3ex}
\begin{quotation}
\floatcaption{tab:diffs}{{\it Differences of $a_\mu^{\rm HVP,lqc}$ values between different ensembles.
All number in units of $10^{-10}$; $a_\mu\equiv a_\mu^{\rm HVP,lqc}$.}}
\end{quotation}
\vspace*{-4.5ex}
\end{table}

Before we do so, we consider the sum of FV, pion-mass mistuning and taste-breaking corrections to all ensembles, and use these to compute the
differences
\begin{equation}
\label{diffs}
a_\mu^{\rm HVP,lqc}(\mbox{ensemble}\ 1)-a_\mu^{\rm HVP,lqc}(\mbox{ensemble}\ 2)\ ,
\end{equation}
both in SChPT and in the SRHO model.   These differences can be compared with data, giving
information on how well SChPT and the SRHO model perform.\footnote{Similar tests were carried out in Ref.~\cite{BMW20}.} The 
differences are shown in Table~\ref{tab:diffs}, where we take for ``ensemble\ 1'' the 96 ensemble, and we vary ``ensemble 2."  While the
lattice numbers have relatively large errors, we see that NNLO SChPT describes these differences reasonably well (possibly thanks to the
large errors on the lattice differences), as does the SRHO model, to a somewhat
lesser extent.\footnote{The agreement is less good for the 48II ensemble.}   The results of Table~\ref{tab:diffs} confirm that between the options shown, NNLO SChPT gives the better description of the data, even on the coarser ensembles.
It is difficult to estimate the systematic errors on the NNLO-SChPT differences in Table~\ref{tab:diffs}, because we do not know
how the errors on the individual contributions in these differences are correlated, and SChPT may not converge for the coarser ensembles.   It is not possible to
estimate a systematic error for the SRHO-based value.

\vskip0.8cm
\section{\label{fullamu} Combination with other contributions}
In this brief section, we combine our value for 
$a_\m^{\rm HVP,lqc}$ with other contributions, taken from the literature, to arrive at a value for $a_\m^{\rm HVP,LO}$ which we believe to be a reliable 
estimate within errors.  In order to do this, we need to add the strange and charm quark contributions, as well as the disconnected parts, QED corrections to order $\a$, and strong isospin breaking (SIB) effects to linear order in $m_u-m_d$.\footnote{Other contributions, such as that from the bottom quark, are small enough that we can ignore them
relative to the size of our error in $a_\m^{\rm HVP,lqc}$.}

We will avoid using any values for these other contributions for which the errors contain any correlations with 
our own result for $a_\m^{\rm HVP,lqc}$.  That excludes using any values from Ref.~\cite{FHM19}, which are based on the
same HISQ ensembles, and, thus also any averages from Ref.~\cite{whitepaper} that include HISQ-based results. 

For the strange-quark plus disconnected contribution, we will take our value from Ref.~\cite{BGMPdisc}.   
In Ref.~\cite{BGMPdisc} it was pointed out that the sum of the strange-quark-connected and disconnected parts
can be determined from the experimental $R$-ratio data, and corrected to yield a value for this sum in the
isospin-symmetric limit of pure QCD without QED.  The analysis was carried out using the $R$-ratio 
compilations of Ref.~\cite{DHMZ} and Ref.~\cite{KNT}, thus leading to two different estimates.  Since these
two estimates are compatible with each other, here we take the average, with the larger of the two errors
and half the difference added in quadrature.   This yields
\begin{equation}
\label{sdisc}
a_\m^{\rm sconn+disc}=39.4(2.1)\times 10^{-10}\ .
\end{equation}
This is in excellent agreement with the value
$39.5(2.9)\times 10^{-10}$ of Ref.~\cite{whitepaper}; it is also in good agreement with the value $38.2(1.8)\times 10^{-10}$ of Ref.~\cite{BMW20}.

For the charm contribution, we use the value obtained by averaging the values of Refs.~\cite{RBC18,ETMc,Mainz19}, 
\begin{equation}
\label{charm}
a_\m^{\rm charm}=14.6(0.7)\times 10^{-10}\ ,
\end{equation}
where we took the larger error, and we avoided using results based on staggered simulations (which, however, are in
good agreement).  We note that the average provided in Ref.~\cite{whitepaper} is the same, but has a smaller error.  In the
combination with our value for $a_\m^{\rm HVP,lqc}$, this makes no difference.

For SIB corrections, we take the weighted average of the results of Ref.~\cite{BMW20}, which found $1.93(1.20)\times 10^{-10}$
using lattice QCD, and Ref.~\cite{JLM21}, which found $3.32(89)\times 10^{-10}$, using ChPT.  Since the errors on these results are
not purely statistical, we will use the larger of the two errors, arriving at 
\begin{equation}
\label{amuSIB}
a_\m^{\rm SIB}=2.82(1.20)\times 10^{-10}\ .
\end{equation}
Adding to this the QED corrections of Ref.~\cite{BMW20} (at present, this work
provides the only complete computation of all contributions to these corrections), we find, to leading order in $\a$ and
$m_u-m_d$:
\begin{equation}
\label{QEDIB}
a_\m^{\rm QED+SIB}=2.82(1.20)\times 10^{-10}-1.45(63)\times 10^{-10}=1.4(1.4)\times 10^{-10}\ .
\end{equation}
We note that the estimate of this contribution provided in Ref.~\cite{whitepaper} is based on estimates of only
some of the many QED plus SIB contributions, and, in particular, did not take into account the strong 
cancellation between connected and disconnected SIB corrections.  

Adding the values in Eqs.~(\ref{amuresult}),~(\ref{sdisc}),~(\ref{charm}) and~(\ref{QEDIB}) and the corresponding errors in quadrature leads to the estimate
\begin{equation}
\label{amutotal}
a_\m^{\rm HVP,LO}=701(14)\times 10^{-10}\ .
\end{equation}
The central value of this result is lower than the average published in Ref.~\cite{whitepaper} and the result of Ref.~\cite{BMW20}, but it is consistent within errors with both.

\section{\label{windows} Window quantities}

We now turn to our results for the window quantities $a_\m^{\rm W1,lqc}$ and $a_\m^{\rm W2,lqc}$, where W1 is the standard window between 0.4 and 1.0 fm, and
W2 is the window betweem 1.5 and 1.9 fm, \seef\ Eq.~(\ref{windowdefs}).   As for $a_\mu^{\rm HVP,lqc}$, we will investigate the continuum limit, and test the applicability of SChPT and the
SRHO model.

\begin{boldmath}
\subsection{\label{standard} Window $0.4-1.0$~fm}
\end{boldmath}

\begin{table}[t]
\centering
\begin{tabular}{|c|c|c|c|c|c|}
\hline
ensemble & 96 & 64 &  48I & 32 & 48II \\
\hline
FV NLO & 0.61 & 0.31 &  0.095 & 0.137 & 0.0129 \\
FV NNLO & 0.73 & 0.36 &  0.108 & 0.170 & 0.021 \\
\hline
\hline
FV SRHO & -1.43 & -0.65 & -0.188 & -0.205 & -0.039\\
\hline
\hline
\hline
retune NLO & -0.069 & -0.52 &  -0.22 & -0.19 & -0.064\\
retune NNLO   & -0.165 & -1.22 &   -0.52 & -0.45 & -0.154 \\
\hline
\hline
retune SRHO & -0.056 & -0.41 & -0.17 & -0.15 & -0.053 \\
\hline
\hline
\hline
t. br. NLO & 0.88  & 3.9  &  8.0 & 11.5 & 11.4 \\
t. br. NNLO  & 2.16 & 10.2& 24.2 & 38.9 & 38.7 \\
\hline
\hline
t. br. SRHO &0.41 &2.1 & 5.8 & 9.9 & 9.9 \\
\hline
\end{tabular}
\vskip3ex
\floatcaption{tab:FVmistbW1}{{\it FV, pion-mass retuning and taste-breaking corrections for $a_\mu^{\rm W1,lqc}$, in units of $10^{-10}$, computed in SChPT and in the SRHO model. 
  To be added to correct lattice results; for details, see text.}}
\end{table}

Again, we begin with showing the FV, pion-mass mistuning and taste-breaking corrections, computed in NLO and NNLO SChPT, as well as in the SRHO model, in Table~\ref{tab:FVmistbW1}.
As we will see in more detail below, SChPT is not reliable for the W1 window, and we thus will not discuss this table in as much detail as we 
did in Sec.~\ref{full} for $a_\m^{\rm HVP,lqc}$.   To the extent that the values in the table can be taken as a guide, we see that corrections are
much smaller for $a_\m^{\rm W1,lqc}$, in comparison with $a_\m^{\rm HVP,lqc}$.  

\begin{figure}[t]
\vspace*{4ex}
\begin{center}
\includegraphics*[width=7cm]{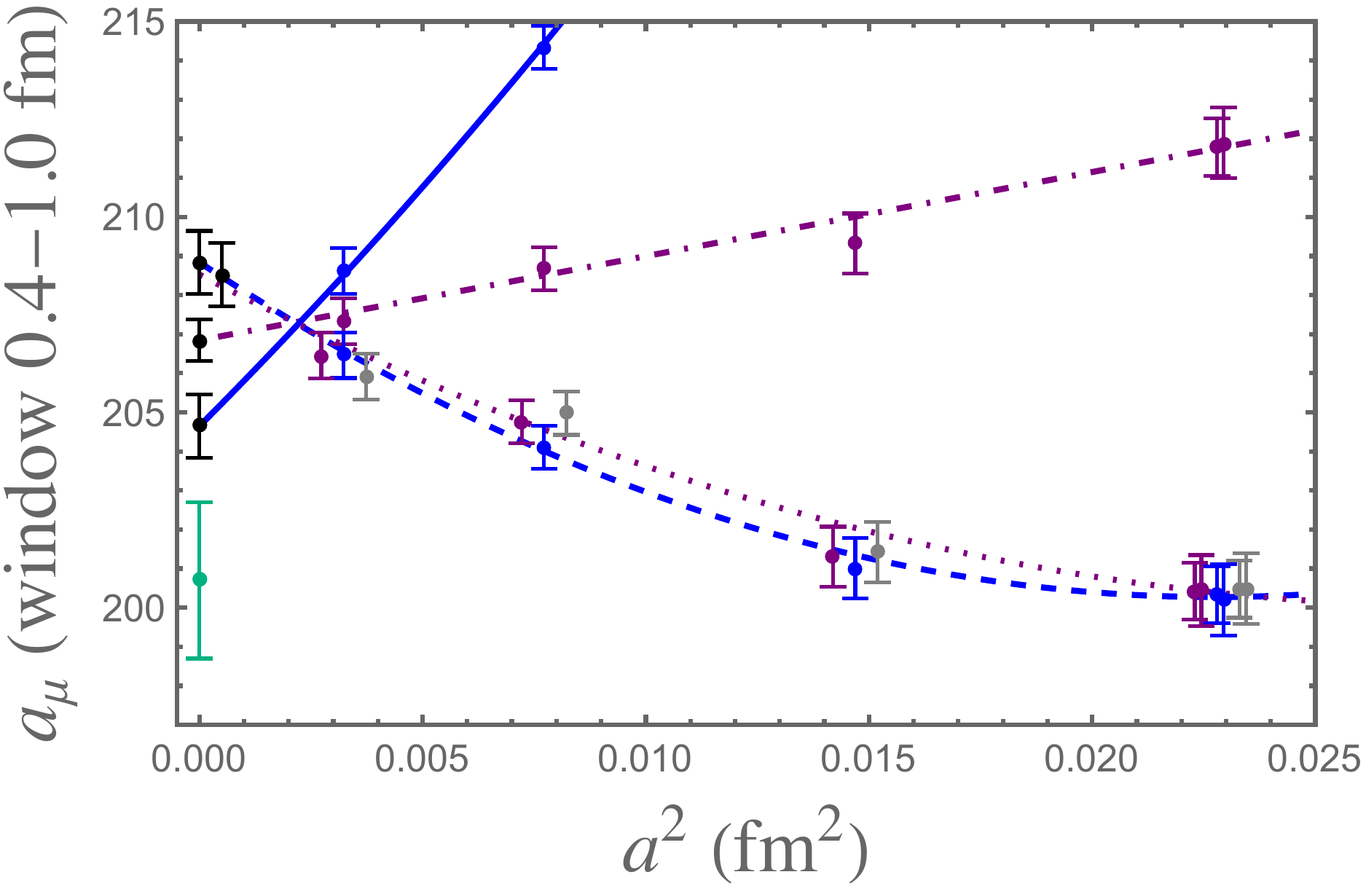}
\hspace{0.5cm}
\includegraphics*[width=7cm]{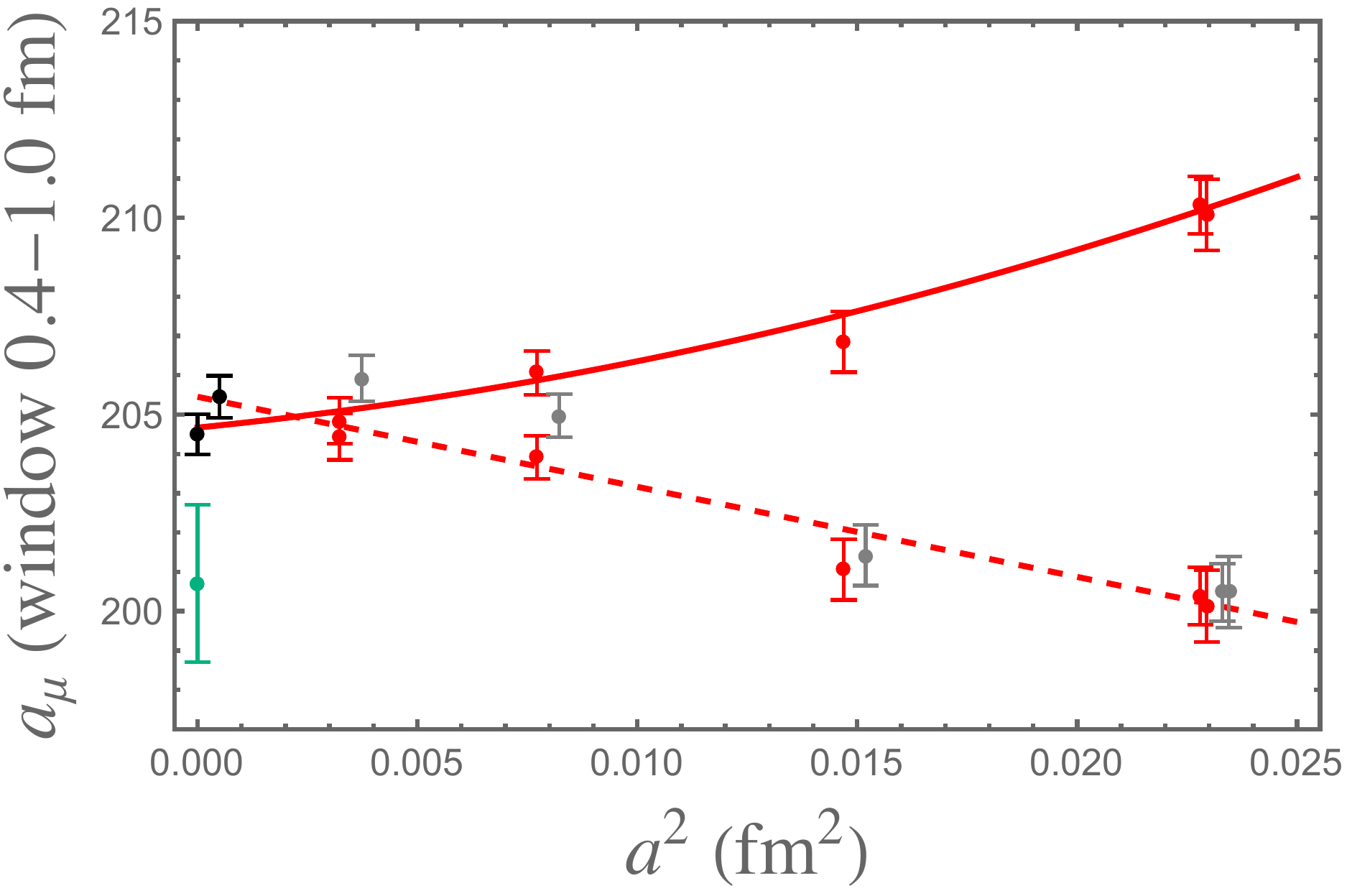}
\end{center}
\begin{quotation}
\floatcaption{fig:window1}%
{{\it The 0.4-1.0~{\rm fm} window, $a_\mu^{\rm W1,lqc}$, in units of $10^{-10}$. Left panel:  Fits using NNLO SChPT for FV corrections and pion mass retuning; without taste breaking  (quadratic, blue dashed line) and with taste breaking (quadratic, blue solid curve); fits using NLO SChPT
for FV corrections and pion mass retuning; without taste breaking  (quadratic, purple dotted line) and with taste breaking (linear, purple dot-dashed line).   Right panel:
Fits using the SRHO model for FV corrections and pion mass retuning; without taste breaking  (linear, red dashed line) and with taste breaking (quadratic, red solid curve).   The data points for each fit are shown in the same color; continuum limits are shown in black.   The grey points in both panels (slightly horizontally offset for clarity) show the uncorrected values reported in Table~\ref{tab:results}.
The isolated (green) point at $a^2=0$ is the estimated value from R-ratio data (by C.~Lehner, using data from 
Ref.~\cite{KNT18}). Some data points and continuum-limit
extrapolations are slightly offset for clarity.}}
\end{quotation}
\vspace*{-4ex}
\end{figure}

In Fig.~\ref{fig:window1} we show $a_\m^{\rm W1,lqc}$ as a function of $a^2$. The gray points again show the data of Table~\ref{tab:results},
with the nearby colored points the same data, but now corrected for FV or pion-mass mistuning using NNLO or NLO SChPT (left panel) and the SRHO model
(right panel).  
The fits to these data points are not completely justified, because it is not clear that SChPT or the SRHO model are the right tools to apply these corrections
for the W1 window.  In practice, FV and mistuning effects appear to be small,
and we obtain good fits ($p$-values 72\%, 22\% and  16\%, respectively).   In the left panel, the solid blue curve represents a quadratic fit to the data now also 
corrected for taste breaking at NNLO, while the dot-dashed purple line shows a linear fit to the data with taste breaking corrected at NLO.  In the right 
panel, the solid red curve shows a fit with the data also corrected for taste breaking using the SRHO model; $p$ values are 73\%, 50\% and 25\%, respectively.
     The 32- and 48II-ensembles points agree within errors.
Our extrapolated values from these fits are as follows:
\begin{eqnarray}
\label{amuW1fits}
a_\m^{\rm W1,lqc}&=& 208.84(81)\times 10^{-10}\ ,\qquad \mbox{NNLO~without~taste~breaking}\ ,\\
&=&204.65(81)\times 10^{-10}\ ,\qquad \mbox{NNLO}\ ,\nonumber\\
&=&208.52(81)\times 10^{-10}\ ,\qquad \mbox{NLO~without~taste~breaking}\ ,\nonumber\\
&=&206.84(53)\times 10^{-10}\ ,\qquad \mbox{NLO~(linear)}\ ,\nonumber\\
&=&205.45(53)\times 10^{-10}\ ,\qquad \mbox{SRHO~without~taste~breaking\ (linear)}\ ,\nonumber\\ 
&=&204.67(81)\times 10^{-10}\ ,\qquad \mbox{SRHO}\ .\nonumber
\end{eqnarray}
We have also carried out linear fits dropping the two data points at $a=0.15$~fm, \ie, dropping the 32 and 48II ensembles.  The continuum-extrapolated
values differ from the central values above by not more than 1.3 times the fit error in Eq.~(\ref{amuW1fits}) for the NNLO-based fits, not more than 0.6 times the fit error for the NLO-based and SRHO-based fits.

Clearly, the different extrapolations shown in Fig.~\ref{fig:window1} do not yield a common continuum limit---the errors of the different extrapolations do not overlap, with the exception of the two SRHO-based extrapolations.   Moreover, it would be misleading to evaluate their level of agreement visually, as the continuum extrapolations are highly correlated.   
If we take the average between the largest and smallest values in Eq.~(\ref{amuW1fits}), we find
\begin{equation}
\label{W1result}
a_\m^{\rm W1,lqc}=(206.75\pm 0.81\pm 2.10)\times 10^{-10}=206.8(2.2)\times 10^{-10}\ ,
\end{equation}
where the first error is the fit error, and the second error is half the difference between the largest and smallest values.  The combined error is 
obtained by quadrature.   We note that the central value is essentially that given by the NLO linear fit, but with a much larger error.
We also note that the estimated error is dominated by systematics. If, based on the discussion 
of Table~\ref{tab:diffsW1} below, we discard the NNLO-based fits, the same procedure would lead
to the estimate $a_\m^{\rm W1,lqc}=(206.60\pm 0.81\pm 1.93)\times 10^{-10}=206.6(2.1)\times 10^{-10}$, fully consistent with Eq.~(\ref{W1result}). If we would take the average of the two SRHO-based extrapolations, we would obtain
$205.1(9)\times 10^{-10}$, which is also consistent with Eq.~(\ref{W1result}), but with a much smaller error.

The data points which are corrected only for FV effects and pion-mass mistuning in Fig.~\ref{fig:window1} lie close to the gray data points, which are the unmodified lattice results shown on Table~\ref{tab:results}.
This suggests that FV and pion-mass mistuning corrections are small, even if there is serious doubt that NNLO SChPT can be trusted to reliably obtain these
corrections.  The 64-ensemble gray point is farthest from the corresponding corrected point because the pion-mass mistuning correction is larger in that case
(\seef\ Table~\ref{tab:ensembles}).  As can be seen in the figure, the fits to these corrected data points (dashed and dotted curves) are non-linear fits.  This is
not unexpected, given the non-linear behavior seen in the taste splittings, as discussed in Sec.~\ref{taste}.  
With this clearly non-linear behavior, it is
not possible to predict the behavior at lattice spacings well below 0.06~fm reliably, and we thus do not believe that the first fit shown in
Eq.~(\ref{amuW1fits}) can be trusted to yield a reliable continuum limit.   While, as we have argued, all methods to compute corrections are based on models,
one notes that all of them lead to a lower continuum limit when taste-breaking corrections are included.   Moreover, NLO SChPT and the SRHO model, when applied to taste-breaking
corrections, appear to linearize the data (NLO SChPT leads to a good linear fit; the SRHO quadratic fit leads to a curvature $1.2\s$ away from zero).
This leads us to believe that the true continuum-limit value of $a_\m^{\rm W1,lqc}$ probably lies below the value obtained in the first line of
Eq.~(\ref{amuW1fits}).  These considerations lead us to the estimate~(\ref{W1result}) as the best value based on our data.  It is clear that lattice results
at smaller lattice spacings would be very helpful in narrowing down this range.

The value obtained in Eq.~(\ref{W1result}) can be compared to the value we obtained from ensembles 96, 64 and 48I in Ref.~\cite{ABGP19}, which is $209.78(96)\times 10^{-10}$.  Our new value 
is lower, but is still $2\s$ higher than the R-ratio-based number, $200.7(2.0)\times 10^{-10}$, shown in Fig.~\ref{fig:window1}.\footnote{If we use the R-ratio-based
value $200.3(1.3)\times 10^{-10}$ from Ref.~\cite{BMW20}, the value in Eq.~(\ref{W1result}) is $2.5\s$ higher.
However, we prefer to use, for comparison, a value that does not use staggered-fermion results.}  The reason for our
larger error is that we now included NNLO-SChPT- and SRHO-based fits, which give a lower value relative to the value not corrected for taste breaking
than the NLO-SChPT-based fit (which we employed exclusively in Ref.~\cite{ABGP19}). This is even though our new lattice results have  much smaller statistical errors. In Ref.~\cite{ABGP19} we also considered the value obtained from extrapolating the
values at the smallest two lattice spacings, which yields $207.7(1.8)\times 10^{-10}$.  Equation~(\ref{W1result}) is
in good agreement with this value.
 
It is interesting to compare the different fits.  The NNLO and SRHO fits lead to the same continuum limit, despite their different nature.  All fits
based on taste-breaking-corrected points have a positive slope as a function of $a^2$, which is what is also seen for domain-wall fermions \cite{RBC18}.
Visually, the
NLO fit looks appealing---with an excellent linear fit.  As we have emphasized above, all three approaches should be 
considered model approaches, as ChPT is not expected to converge for this window.  In fact, we can look into this by considering the differences~(\ref{diffs}), 
but now for $a_\mu^{\rm W1,lqc}$.

These differences are shown in Table~\ref{tab:diffsW1}.   Clearly, NNLO SChPT does not describe these differences, and the SRHO model does much better.
However, NLO SChPT describes the data as well as the SRHO model.   Combined with our finding that the NLO-corrected data allow for a 
very good linear fit, it would not be unreasonable to conclude that, at the level of a model, NLO SChPT might provide the preferred model.  We do not believe
this to be justified, but instead, we conclude that it is not justified to use a model to correct data points based on visual
improvement alone.  We note that this is reflected in a comparison between our ``best'' result, Eq.~(\ref{W1result}), and the NLO-linear fit in Eq.~(\ref{amuW1fits}),
which lead to the same central value, but very different errors.

In summary, even though from Fig.~\ref{fig:window1} one could infer that the SRHO model should be preferred over
SChPT, for both SChPT and the SRHO model there is no path to a systematic improvement of the model-description of the
data.   In our view, that makes the window W1 of limited usefulness until the removal of systematic effects can be carried out using lattice data,
without the need for any model. Since taste-breaking effects appear to be the largest effect hindering a straightforward extrapolation to the
continuum limit, data at a smaller lattice spacing would go a long way to improving this unsatisfactory situation.

We compare our new value for $a_\m^{\rm W1,lqc}$ with values obtained by other collaborations in Fig.~\ref{fig:window}.
We observe that our new value is consistent with other staggered determinations (first five values in the figure).
In particular, our value is in agreement with that obtained in Ref.~\cite{BMW20}, even though we assigned a larger
systematic error, \seef\ Eq.~(\ref{W1result}).   Comparing the right-hand panel in Fig.~\ref{fig:window1} with Fig.~4
of Ref.~\cite{BMW20}, one notices the similarity between these two figures.\footnote{The comparison should be made
between the data points shown in red in Fig.~\ref{fig:window1} and the data points in Fig.~4 of Ref.~\cite{BMW20},
because the ``no improvement'' data points in that figure have already been extrapolated to infinite volume.
We note that Ref.~\cite{BMW20} and we both used conserved currents.}
Data points without corrections for taste breaking span about the same range of values for $a_\m^{\rm W1,lqc}$,
and this is also true for the taste-breaking corrected points.  In this comparison, we compare the 
lattice spacings of Ref.~\cite{BMW20} directly to ours, even though Ref.~\cite{BMW20} used a different,
but also highly improved, staggered action.  This appears to be justified by the observation that taste-breaking
effects, as modeled by the SRHO model, are of approximately the same size at the same lattice spacing in both
Ref.~\cite{BMW20} and this work.  This suggests that also with the action of Ref.~\cite{BMW20}, it would be
desirable to see what happens at a smaller lattice spacing. 

A similar comparison can be made between Fig.~\ref{amu} and the Extended Data Fig.~3 of Ref.~\cite{BMW20}, for the
data without corrections for taste breaking (the green triangles in Ref.~\cite{BMW20}).  As for $a_\m^{\rm W1,lqc}$,
they span about the same range, confirming that lattice spacings can be directly compared between Ref.~\cite{BMW20}
and this paper, and taste-breaking effects are comparable in size.  

\begin{table}[t]
\begin{center}
\begin{tabular}{|c|c|c|c|c|}
\hline
 & ${\rm W1}(96)-{\rm W1}(64)$ & ${\rm W1}(96)-{\rm W1}$(48I) & ${\rm W1}(96)-{\rm W1}(32)$ & ${\rm W1}(96)-{\rm W1}$(48II)\\
\hline
lattice & 0.94(46) & 4.49(66) & 5.43(79) & 5.44(58) \\
\hline
NLO~SChPT & 2.28 & 6.47 & 9.98 & 9.89\\
NNLO~SChPT & 6.67 & 21.08 & 35.88 & 35.87 \\
\hline
SRHO & 2.15 & 6.49 & 10.65 & 10.91 \\
\hline
\end{tabular}
\end{center}
\vspace*{-3ex}
\begin{quotation}
\floatcaption{tab:diffsW1}{{\it Differences of $a_\mu^{\rm W1,lqc}$ values between different ensembles.
All number in units of $10^{-10}$; ${\rm W1}\equiv a_\mu^{\rm W1,lqc}$.}}
\end{quotation}
\vspace*{-4.5ex}
\end{table}

\begin{boldmath}
\subsection{\label{new} Window $1.5-1.9$~fm}
\end{boldmath}
We now consider our new window, W2.  Since 1.5~fm$\,\approx\,$(130~MeV)$^{-1}$ is a rather large distance, it is reasonable to expect that ChPT can be used to describe
$a_\mu^{\rm W2,lqc}$.   In Table~\ref{tab:FVmistbW2} we show the FV, pion-mass mistuning and taste-breaking corrections computed in SChPT and in the SHRO model, in the same
format as in Table~\ref{tab:FVmistbW1}.  We note that the convergence of SChPT for window W2 is much better in 
general than for window W1, even though taste-breaking corrections at NNLO are still more than 100\% larger than at
NLO for the 48I, 32 and 48II ensembles.

\begin{table}[t]
\vspace*{4ex}
\centering
\begin{tabular}{|c|c|c|c|c|c|}
\hline
ensemble & 96 & 64 &  48I & 32 & 48II \\
\hline
FV NLO & 2.25 & 1.08 & 0.325 & 0.372 & 0.062 \\
FV NNLO & 3.13 & 1.51 & 0.448 & 0.559 & 0.0744 \\
\hline
\hline
FV SRHO & 0.56 & 0.35 & 0.128 &  0.197 & 0.0156\\
\hline
\hline
\hline
retune NLO & -0.114 & -0.879 &  -0.364 & -0.313 & -0.107\\
retune NNLO & -0.199 & -1.52  & -0.631 & -0.543 & -0.186 \\
\hline
\hline
retune SRHO & -0.280 & -2.12 & -0.885 & -0.761 & -0.262 \\
\hline
\hline
\hline
t. br. NLO & 1.39 & 5.42 &  8.55 & 10.12 & 9.94 \\
t. br. NNLO & 2.47 & 10.14 &  17.60 & 22.06 & 21.76 \\
\hline
\hline
t. br. SRHO & 3.01 & 12.84 & 24.54 & 33.81 & 33.46 \\
\hline
\end{tabular}
\vskip3ex
\floatcaption{tab:FVmistbW2}{{\it FV, pion-mass retuning and taste-breaking corrections for $a_\mu^{\rm W2,lqc}$, in units of $10^{-10}$, computed in SChPT and in the SRHO model. 
  To be added to correct lattice results; for details, see text.}}
\end{table}

\begin{figure}[t]
\vspace*{4ex}
\begin{center}
\includegraphics*[width=7cm]{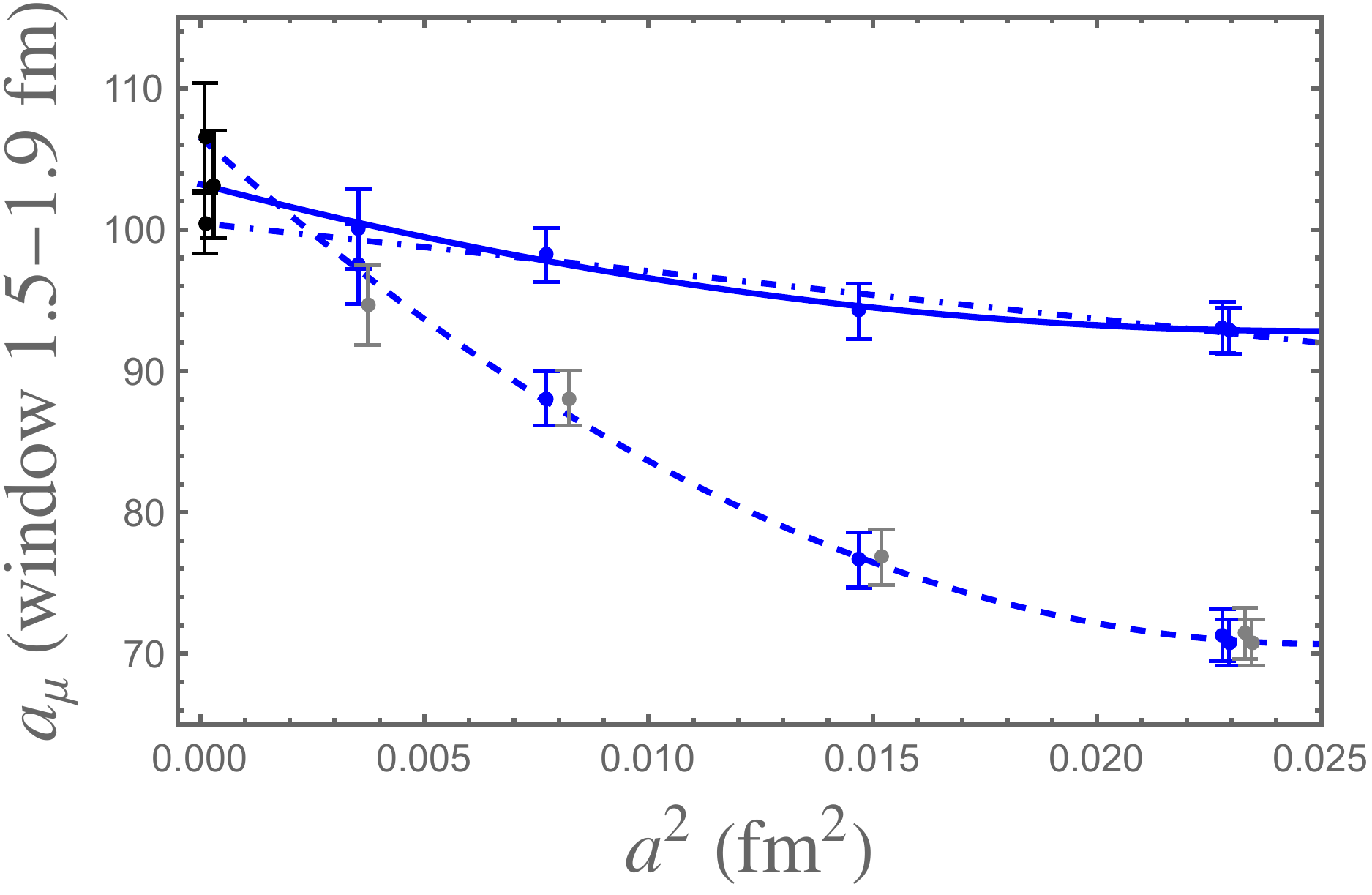}
\hspace{0.5cm}
\includegraphics*[width=7cm]{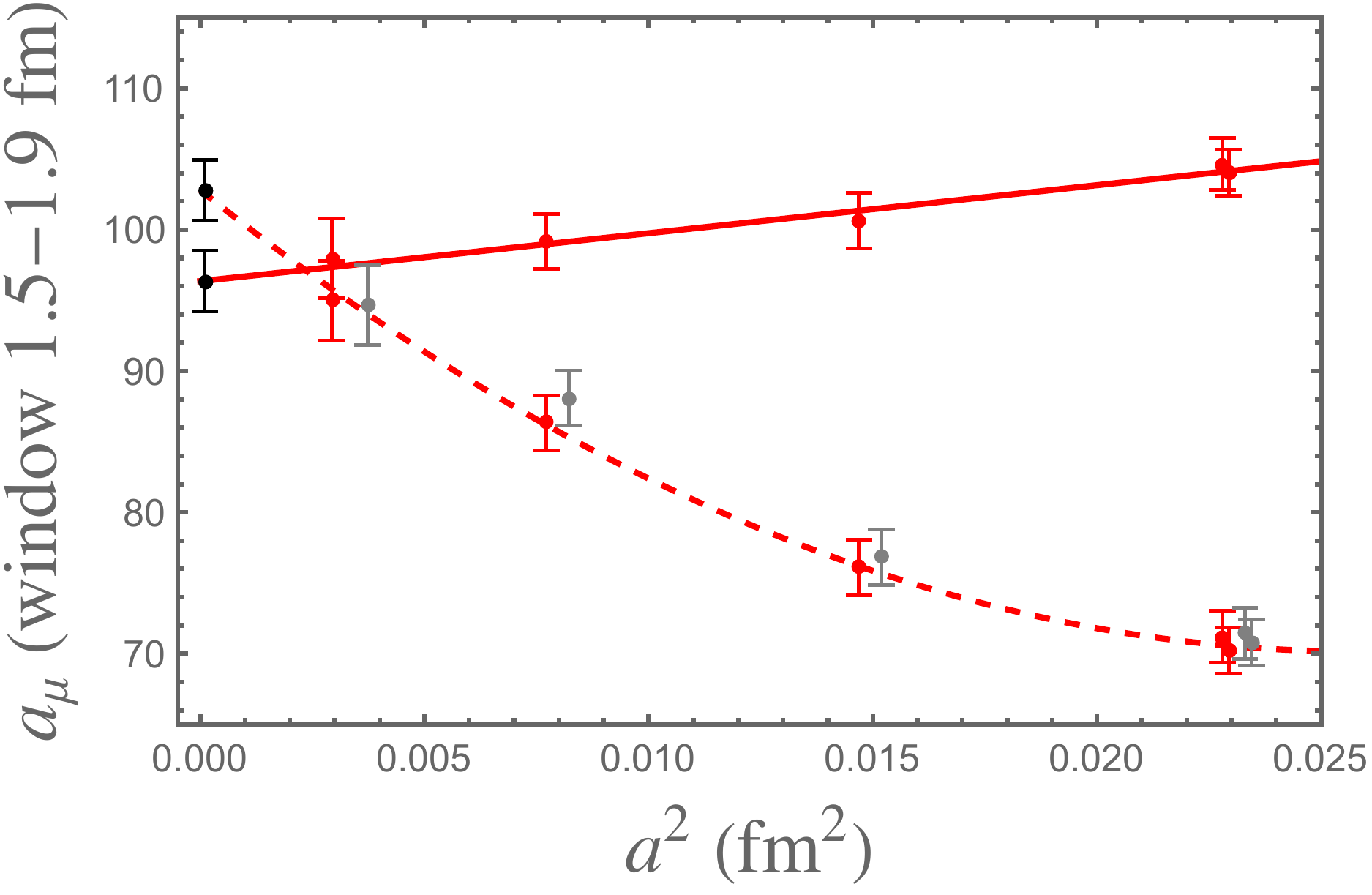}
\end{center}
\vspace*{-4ex}
\begin{quotation}
\floatcaption{fig:window2}%
{{\it The 1.5-1.9~{\rm fm} window, $a_\mu^{\rm W2,lqc}$, in units of $10^{-10}$. Left panel:  Fits using NNLO SChPT for FV corrections and pion mass retuning; without taste breaking  (quadratic, blue dashed line) and with taste breaking (quadratic, blue solid curve; linear, dot-dashed line).   Right panel:
Fits using the SRHO model for FV corrections and pion mass retuning; without taste breaking  (quadratic, red dashed line) and with taste breaking (linear, red solid curve).   The data points for each fit are shown in the same color; continuum limits are shown in black.   The grey points in both panels (slightly horizontally offset for clarity) show the uncorrected values reported in Table~\ref{tab:results}. Some data points and continuum-limit
extrapolations are slightly offset for clarity.}}
\end{quotation}
\end{figure}
In Fig.~\ref{fig:window2} we show the lattice spacing dependence of $a_\m^{\rm W2,lqc}$.    The gray points again show the data of Table~\ref{tab:results},
with the nearby colored points the same data, but now corrected for FV or pion-mass mistuning using NNLO SChPT (left panel) and the SRHO model
(right panel).   All fits shown in the figure have excellent $p$-values,
which reflects the fact that the statistical errors on the W2 values are relatively large. 
We note that the NNLO taste-breaking corrections appear to capture a large
fraction of the lattice spacing effects for window W2, witness the relatively small slope of the blue
solid and dot-dashed curves in the left-hand panel.
Our extrapolated values from these fits are as follows:
\begin{eqnarray}
\label{amuW2fits}
a_\m^{\rm W2,lqc}&=& 106.5(3.8)\times 10^{-10}\ ,\qquad \mbox{NNLO~without~taste~breaking}\ ,\\
&=&103.2(3.8)\times 10^{-10}\ ,\qquad \mbox{NNLO}\ ,\nonumber\\
&=&100.5(2.1)\times 10^{-10}\ ,\qquad \mbox{NNLO\ (linear)}\ ,\nonumber\\
&=&102.8(3.8)\times 10^{-10}\ ,\qquad \mbox{SRHO~without~taste~breaking}\ .\nonumber\\
&=&\phantom{1}96.4(2.1)\times 10^{-10}\ ,\qquad \mbox{SRHO~(linear)}\ .\nonumber
\end{eqnarray}
In this case, the continuum limit obtained shows better agreement between the values 
with and without taste-breaking corrections when these are computed with NNLO SChPT, than when 
they are computed with the SRHO model.  If we interpret this as an indication that NNLO SChPT 
is more reliable, averaging the largest and smallest NNLO-based values in Eq.~(\ref{amuW2fits}) 
would give an estimated best value of $a_\m^{\rm W2,lqc}=103(4)\times 10^{-10}$, where we 
averaged the first and third values in Eq.~(\ref{amuW2fits}), and took the largest fit error as
our error estimate.  The SRHO model would lead to a lower estimate, which, based on our 
discussion of Table~\ref{tab:diffsW2} (see below), cannot be excluded.
We have also carried out linear fits dropping the two data points at $a=0.15$~fm, \ie, dropping the 32 and 48II ensembles.  The continuum-extrapolated
values differ from the central values above by not more than 1.2 times the fit errors in Eq.~(\ref{amuW2fits}) for the NNLO case.
In this case, the continuum values obtained with NNLO SChPT with and without taste breaking are equal to
$102.1(2.4)\times 10^{-10}$ in both cases, and agree in the continuum limit.  We thus take 
\begin{equation}
\label{W2result}
a_\m^{\rm W2,lqc}=102.1(2.4)\times 10^{-10}\ ,
\end{equation}
as our best estimate for window W2.  This result is
in good agreement with the value $103(4)\times 10^{-10}$ quoted above. 

\begin{table}[t]
\begin{center}
\begin{tabular}{|c|c|c|c|c|}
\hline
 & ${\rm W2}(96)-{\rm W2}(64)$ & ${\rm W2}(96)-{\rm W2}(48\mbox{I})$ & ${\rm W2}(96)-{\rm W2}(32)$ & ${\rm W2}(96)-{\rm W2}(48\mbox{II})$ \\
\hline
lattice & 6.6(2.7) & 17.8(2.8) & 23.9(2.6) & 23.2(2.7) \\
\hline
NLO~SChPT & 2.1 & 5.0 & 6.7 & 6.4 \\
NNLO~SChPT & 4.7 & 12.0 & 16.7 & 16.3  \\
\hline
SRHO & 7.8 & 20.5 & 30.0 & 29.9 \\
\hline
\end{tabular}
\end{center}
\vspace*{-3ex}
\begin{quotation}
\floatcaption{tab:diffsW2}{{\it Differences of $a_\mu^{\rm W2,lqc}$ values between different ensembles.
All number in units of $10^{-10}$; ${\rm W2}\equiv a_\mu^{\rm W2,lqc}$.}}
\end{quotation}
\vspace*{-4.5ex}
\end{table}
Differences as defined in Eq.~(\ref{diffs}) for the W2 window are tabulated in Table~\ref{tab:diffsW2}.  In this case, NNLO SChPT describes the differences much better than for the W1 window, and also better than NLO SChPT.   The change from NLO to NNLO SChPT is large, but, as remarked before, this does not necessarily mean that ChPT does not converge. We also observe that the agreement between the lattice and NNLO-SChPT numbers is better for ensembles with a smaller lattice spacing: within errors, the lattice difference in column 2 agree with NNLO SChPT, with the tension gradually increasing in columns 3 and 4, \ie, with increasing taste masses.  As for \amuHVP, it is possible that some of the staggered pion masses on the 48I and 32 ensembles are too large, \seef\ Sec.~\ref{comp}.  All this makes it 
difficult to estimate the systematic error from truncating SChPT at NNLO, and it is thus not straightforward to assess the agreement with the lattice values for these differences.  The SRHO model also describes the lattice differences reasonably well, but, of course, it is not possible to assess the systematic error at all.  
Smaller errors on the lattice differences would help discriminate between NNLO SChPT and the SRHO model.
A reduction of the errors on $a_\m^{\rm W2,lqc}$ shown in Table~\ref{tab:results} and/or smaller lattice spacings will be needed to obtain a more precise 
estimate for $a_\m^{\rm W2,lqc}$.

\section{\label{concl} Conclusion}

In this paper, we continued our study of the light-quark connected contribution $a_\m^{\rm HVP,lqc}$
to the HVP part of the
muon anomalous magnetic moment, using lattice QCD with staggered fermions.  We presented and discussed
results for $a_\m^{\rm HVP,lqc}$, the ``intermediate-distance'' window quantity $a_\m^{\rm W1,lqc}$ of Ref.~\cite{RBC18}, and a
new window quantity $a_\m^{\rm W2,lqc}$ that probes the region between 1.5 and 1.9~fm.   We extended our use of 
SChPT to NNLO, in order to compute FV, pion-mass mistuning, and taste-breaking corrections.  For 
comparison, we also computed these corrections using the SRHO model of Ref.~\cite{HPQCD16}.

We now have values for each of these quantities at four different lattice spacings, adding two
ensembles at $a=0.15$~fm, and improving statistics on the $96$, $64$ and 48I ensembles. In
particular for the 96 ensemble, statistical errors are significantly smaller than in Ref.~\cite{ABGP19}.
This allows for a 
more detailed study of the continuum limit than was possible in Ref.~\cite{ABGP19}.  In order to do this,
FV and pion-mass retunings need to be estimated on all ensembles, because none have precisely the
same (spatial) volume and pion mass (even though the pion masses are close to physical).   Taste-breaking
corrections can also be computed, but, with four different lattice spacings, direct extrapolation is
also possible, in principle at least.

This raises the important question how to compute these corrections.  As we emphasized in Ref.~\cite{ABGPEFT}, $a_\m^{\rm HVP,lqc}$ is accessible to ChPT extended to include muons and photons, and, to NNLO, no new counter terms are needed beyond those already present for the strong interaction only.
Here we showed that NNLO ChPT (in infinite volume) can be expected to be reliable for pion masses
up to roughly 250~MeV, based on a comparison with Ref.~\cite{Coletal}.   In addition, the largest 
uncertainty comes from the contribution of the $\co(p^6)$ LEC $c_{56}$, which is independent of
the volume and pion masses, so it drops out of differences.   At the same time, on our coarser
ensembles some of the taste masses are significantly larger than 250~MeV, making it less clear
that one can rely on NNLO SChPT.  This is one argument for going to lattice spacings smaller
than the currently smallest lattice spacing, $a\approx 0.057$~fm.  

Our new value for $a_\m^{\rm HVP,lqc}$, Eq.~(\ref{amuresult}) is smaller, and has a smaller
total error, than the value we obtained in Ref.~\cite{ABGP19}.  This is primarily explained by the fact 
that our uncorrected lattice value for the 96 ensemble is now lower, by 
$17\times 10^{-10}$, even if it is consistent with our previous result within errors.  This difference is to be compared
with the difference of $13\times 10^{-10}$ between our value for $a_\m^{\rm HVP,lqc}$ in 
Ref.~\cite{ABGP19} and our best new value, obtained in Eq.~(\ref{amuresult}).  It is also interesting to
compare the reduction in the total error on our final result, from $22\times 10^{-10}$ 
in Ref.~\cite{ABGP19} to $14\times 10^{-10}$ in Eq.~(\ref{amuresult}),
to the reduction of the error on the uncorrected value on the 96 ensemble,
from $27.5\times 10^{-10}$ to $13\times 10^{-10}$.  While the use of NNLO SChPT has made the computation of
the various corrections that need to be applied to the uncorrected lattice results more precise, it is clear
that the improvement of statistics on our 96 ensemble plays a major role in the reduction of the overall error.
Unfortunately,
our error on $a_\m^{\rm HVP,lqc}$ is large enough for our value 
to agree both with the value obtained in
Ref.~\cite{BMW20}, and with the value obtained from the data-driven 
approach \cite{whitepaper}, if we assume that the discrepancy between these two comes from the light-quark connected part only.  We combined our value for $a_\m^{\rm HVP,lqc}$ with the contributions from strange, charm, disconnected, QED
and strong isospin breaking into an estimate for $a_\m^{\rm HVP,LO}$ in Eq.~(\ref{amutotal}).

The window quantity $a_\m^{\rm W1,lqc}$ covers the 0.4--1.0~fm range, and is not accessible to
SChPT, in agreement with the conclusion reached in Ref.~\cite{BMW20}.   This presents us with 
a conundrum:  the comparison of values for $a_\m^{\rm W1,lqc}$ between different lattice
collaborations is only useful if the FV effects, pion-mass retuning, and, in the staggered case,
taste breaking, can be reliably controlled.  While in principle this can be done numerically,
on the lattice, by considering very large volumes and very small lattice spacings, the present
state of the art does not allow us to do this, and this forces one to rely on a model to
compute all relevant corrections.   Here, we explored three parametrizations: NLO and NNLO SChPT
and the SRHO model, which all should be considered models for the case of the W1 window.   
Clearly, NNLO SChPT does extremely poorly, but, judged by comparisons,
NLO SChPT and the SRHO model do about equally well.  In our case, based on the comparison
of NLO SChPT and the SRHO model in Eq.~(\ref{amuW1fits}), this leads to a large systematic error,
which diminishes the advantage that $a_\m^{\rm W1,lqc}$ can be computed with very small
statistical errors.  Simulations at smaller lattice spacings (with small statistical
errors) would help reduce this systematic error.  As we discussed at the end of Sec.~\ref{standard},
our data for the W1 window look very similar to those of Ref.~\cite{BMW20}, which suggests that
our conclusions are relevant for simulations based on the action of Ref.~\cite{BMW20} as well.

Because of these issues with window W1, we have also investigated a new window, W2, which
covers the 1.5--1.9~fm range.  The advantage of this window is that it can reasonably be
expected to be accessible to ChPT, while the disadvantage is that statistical errors will
be larger.  Indeed, we find that statistical errors are larger, about 3--6 times as large as
for window W1, but still about a factor 5 times smaller than those on $a_\m^{\rm HVP,lqc}$.
Table~\ref{tab:diffsW2} suggests that indeed NNLO SChPT provides a reasonable description
of this window, in sharp contrast with Table~\ref{tab:diffsW1} for window W1.  However,
given the large taste pion masses on ensembles 48I, 32 and 48II, for which SChPT may not
converge, again simulations at a smaller lattice spacing should provide more insight.
While our results for the W2 window have larger errors than our results for the W1
window, we believe it is important
to consider ``auxiliary'' quantities within the domain of validity of ChPT in the future.   

We believe it is fair to say that taste-breaking effects in $a_\m^{\rm HVP,lqc}$ and the window
quantities on these ensembles are not well understood.
This is reflected by the fact that
lattice results uncorrected for taste breaking do not generally extrapolate to the same 
continuum limit as those that are corrected for taste breaking, despite the fact that taste
breaking is a pure lattice artifact.  While the differences in the continuum limit can be
accounted for as a systematic error, it is not clear that this is sufficient for complete
control of the continuum limit. It may well be that the current discrepancies of 
lattice results for $a_\m^{\rm W1,lqc}$ with values obtained from the
dispersive approach, observed in a number of simulations \cite{BMW20,ABGP19,LM20,FHM20,chiQCD22} and this work,
are due to short-distance effects caused by the lack of a sufficiently detailed understanding of the
continuum limit.  As this discrepancy for the intermediate window W1 is of order half the
discrepancy between the lattice value for \amuHVP\ of Ref.~\cite{BMW20} and the dispersive value, this 
 suggests that the discrepancy in \amuHVP\ itself may be caused by the same
short-distance effects. Of course, it is just this type of focus that the windows were designed to facilitate.

In fact, it appears that taste breaking on these HISQ ensembles itself is not well understood.
SChPT predicts an approximate $SO(4)$ symmetry at order $a^2$ \cite{LS}, and indeed, the
taste splittings shown in Fig.~\ref{tastespectrum} exhibit this $SO(4)$ symmetry.   However, at the same
time, we find that the coefficient of the $a^2\a_s^2$ term in the fit is consistent with
zero.  Since at order $a^4$ operators appear in SChPT that break the continuum $SU(4)$ taste
symmetry down to the minimal lattice symmetry group \cite{MGmesons}, this would imply that the HISQ action
somehow suppresses those operators at order $a^4$ that break $SO(4)$ down to the lattice
symmetry group.  With the present set of ensembles, taste splittings are quite non-linear
as a function of $a^2$, and this is reflected in the generally non-linear behavior seen in
Figs.~\ref{fig:window1} and \ref{fig:window2}.

Returning to $a_\m^{\rm HVP,lqc}$ and the window quantities,
in order to control taste breaking more reliably, it would be very helpful 
to reach the regime in which taste breaking is approximately linear in $a^2\a_s^2(1/a)$.   
With the HISQ action, this means going to smaller lattice spacings.   In addition, taste
splittings have to be small enough (of order $\ltap\,150$~MeV) for NNLO SChPT to be 
applicable.  Finally, there are likely to be scaling violations from other sources than
taste breaking, and one should thus not necessarily expect the data points corrected for taste
breaking to become constant as a function of the lattice spacing.

We summarize our main conclusions. First, in order to reliably extrapolate lattice
results obtained with staggered fermions to the continuum limit, smaller lattice spacings
will be needed, at least with the HISQ action.  
Adding at least one smaller lattice spacing would allow us to dispense
with the 48I,II and 32 ensembles in taking the continuum limit.   Furthermore, as can be
seen in Table~\ref{tab:results}, the statistical errors on our lattice results are now
competitive with scale-setting errors, especially for the window quantities.  
The total error can thus be reduced by a more
precise determination of the scale.
Second, as long as it will be
necessary to apply corrections for finite volume, pion mass retuning, and, in the
case of staggered fermions, taste-breaking, it is crucial to consider quantities for
which these corrections can be reliably computed using effective field theory methods,
while models, in a ``first-principles'' computation, should be avoided. 

\vspace{3ex}
\noindent {\bf Acknowledgments}
\vspace{3ex}

We thank Claude Bernard for discussions about taste breaking, 
Kim Maltman for discussions about isospin breaking,
Doug Toussaint (for MILC) for providing the full taste-pion spectrum on the
96, 64, 48I and 32 ensembles, the MILC collaboration for the use of their gauge configurations,
and Andr\'e Walker-Loud (for CalLat) for making the 48II ensemble available to us.
This work used the Extreme Science and Engineering Discovery Environment (XSEDE),
which is supported by National Science Foundation grant number ACI-1548562.
We thank the Pittsburgh Supercomputing Center
(PSC), the San Diego Supercomputer Center (SDSC), and  the Texas Advanced Computing
Center (TACC), 
where the lattice computations were performed. 
TB and MG are supported by the U.S.\ Department 
of Energy, Office of Science, Office of High Energy Physics, under 
Awards DE-SC0010339 and DE-SC0013682, respectively. SP is supported 
by the Spanish Ministry of Science, Innovation and Universities 
(project PID2020-112965GB-I00/AEI/10.13039/501100011033) and by 
Grant 2017 SGR 1069. IFAE is partially funded by the CERCA 
program of the Generalitat de Catalunya.

\appendix
\section{\label{SRHO} SRHO model}
We use the SRHO model, as first introduced in Ref.~\cite{HPQCD16}, in the implementation of Ref.~\cite{BMW20}.  We chose
\begin{eqnarray}
\label{SRHpars}
m_\r&=& 775~\mbox{MeV}\ ,\\
F_\r&=& 210~\mbox{MeV}\ ,\nonumber\\
g_\r&=&6\ ,\nonumber\\
g_\g&=&5.4\ .\nonumber
\end{eqnarray}
In infinite volume, it is straightforward to obtain the spectral function 
$\r_{\rm SRHO}(s)$ from the SRHO-model version of the vacuum polarization, and, from this,
\begin{equation}
\label{Ctfromrho}
C_{\rm SRHO}(t)=\frac{10}{9}\,\half\int_{4m_\p^2}^\infty ds\,\r_{\rm SRHO}(s)\sqrt{s}\,e^{-\sqrt{s}t}\ ,
\end{equation}
where the factor 10/9 is needed to obtain the quark-connected part.
This gives us access to the SRHO-model prediction for $a_\m^{\rm HVP,lqc}$
and $a_\m^{\rm W1/2,lqc}$, as a function of the pion mass, which allows us
to get SRHO-model predictions for pion-mass retuning and taste breaking
in infinite volume.   For FV corrections in the SRHO model, we use 
Poisson resummation, combined with the strategy of Ref.~\cite{BMW20} to incorporate the window in momentum space through the introduction of the window version
$\hat{\P}_{\rm win}(Q^2)$ of $\hat{\P}(Q^2)$.


\end{document}